\documentclass[%
reprint,
superscriptaddress,
amsmath,amssymb,
aps,
prb,
]{revtex4-2}

\usepackage{graphicx}% Include figure files
\usepackage{dcolumn}% Align table columns on decimal point
\usepackage{bm}% bold math
\usepackage{braket}
\usepackage{appendix}
\usepackage{subfigure}
\usepackage{hyperref}
\usepackage{color}
\usepackage{xcolor}
\usepackage{amsmath}
\usepackage{nccmath}
\usepackage[english]{babel}
\usepackage{svg}

\begin{document}

	\title{Quantum sensing of paramagnetic spins in liquids with spin qubits in \\ hexagonal boron nitride}% Force line breaks with \\
	\author{Xingyu Gao}
	\affiliation{
		Department of Physics and Astronomy, Purdue University, West Lafayette, Indiana 47907, USA
	}
	\author{Sumukh Vaidya}
	\affiliation{
		Department of Physics and Astronomy, Purdue University, West Lafayette, Indiana 47907, USA
	}
	\author{Peng Ju}
	\affiliation{
		Department of Physics and Astronomy, Purdue University, West Lafayette, Indiana 47907, USA
	}
		\author{Saakshi Dikshit}%
	\affiliation{%
		Elmore Family School of Electrical and Computer Engineering, Purdue University, West Lafayette, Indiana 47907, USA
	}%
	
	\author{Kunhong Shen}%
	\affiliation{
		Department of Physics and Astronomy, Purdue University, West Lafayette, Indiana 47907, USA
	}
	\author{Yong P. Chen}%
	\affiliation{
		Department of Physics and Astronomy, Purdue University, West Lafayette, Indiana 47907, USA
	}
	\affiliation{%
		Elmore Family School of Electrical and Computer Engineering, Purdue University, West Lafayette, Indiana 47907, USA
	}
	\affiliation{
		Purdue Quantum Science and Engineering Institute, Purdue University, West Lafayette, Indiana 47907, USA
	}
	\affiliation{
		Birck Nanotechnology Center, Purdue University, West Lafayette, Indiana 47907, USA
	}
	\author{Tongcang Li}%
	\email{tcli@purdue.edu}
	\affiliation{
		Department of Physics and Astronomy, Purdue University, West Lafayette, Indiana 47907, USA
	}
	\affiliation{%
		Elmore Family School of Electrical and Computer Engineering, Purdue University, West Lafayette, Indiana 47907, USA
	}
	\affiliation{
		Purdue Quantum Science and Engineering Institute, Purdue University, West Lafayette, Indiana 47907, USA
	}
	\affiliation{
		Birck Nanotechnology Center, Purdue University, West Lafayette, Indiana 47907, USA
	}
	\date{\today}% It is always \today, today,

\begin{abstract}
Paramagnetic ions and radicals play essential roles in biology and medicine, but detecting these species requires a highly sensitive and ambient-operable sensor. Optically addressable spin color centers in 3D semiconductors have been used for detecting  paramagnetic spins as they are sensitive to the spin magnetic noise.  However, the distance between spin color centers and target spins is limited due to the difficulty of creating high-quality spin defects near the surface of 3D materials. Here, we show that spin qubits in hexagonal boron nitride (hBN), a layered van der Waals (vdW) material, can serve as a promising sensor for nanoscale detection of paramagnetic spins in liquids. We first create shallow spin defects in close proximity to the hBN surface, which sustain high-contrast optically detected magnetic resonance (ODMR) in liquids at room temperature. Then we demonstrate sensing  spin noise of paramagnetic ions in water based on spin relaxation measurements. Finally, we show that paramagnetic ions can reduce the contrast of spin-dependent fluorescence, enabling efficient detection by continuous wave ODMR. Our results demonstrate the potential of ultrathin hBN quantum sensors for chemical and biological applications.

\end{abstract}

\maketitle

%\textbf {Keywords:} {spin defects, hexagonal boron nitride, quantum sensing, optically detected magnetic resonance, paramagnetic ions}

%\section{Main}

\begin{figure*}[t]
	\centering
	\includegraphics[width=1\textwidth]{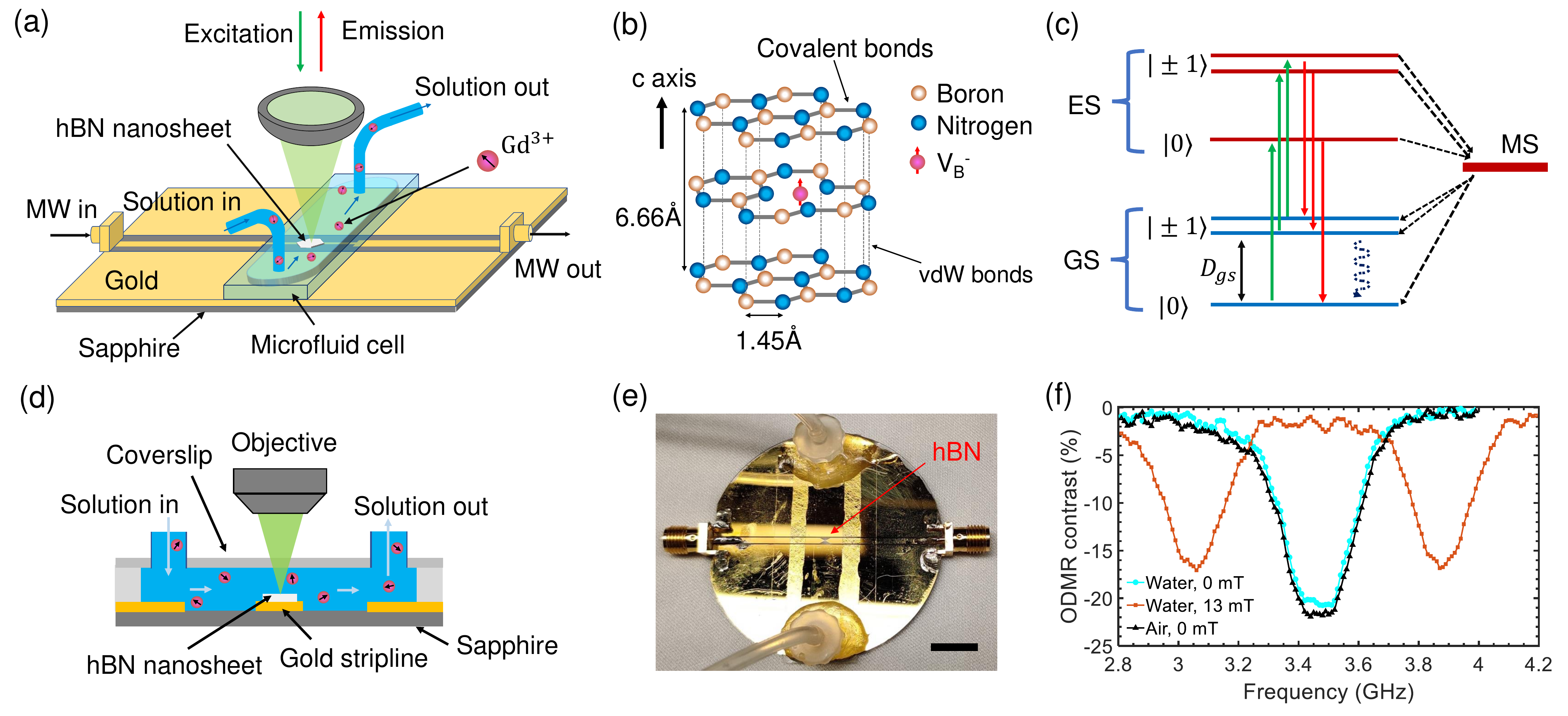}
	\caption{ \textbf{In-solution quantum sensing with shallow spin defects in hBN.} (a) Schematic of the experimental setup. An hBN nanosheet with $V_B^-$ spin defects is transferred onto a gold  stripline microwave (MW) waveguide on a sapphire substrate. A glass coverslip is placed on top of the waveguide and spaced by double-side tapes. Two tubes are connected to  two sides of the coverslip for delivering and switching solutions. The device is sealed with epoxy. An objective lens focuses a 532 nm green laser to excite $V_B^-$ defect spins, and collects the emission from $V_B^-$ defect spins.  A microwave is delivered by the waveguide to drive spin defects.  (b) Atomic structure of a $V_B^-$  defect. The c-axis is perpendicular to the hBN 2D lattice. (c) Simplified energy levels of a $V_B^-$ defect, which include a triplet (S = 1) ground state, a triplet (S = 1) excited state, and a singlet metastable state. (d) Schematic of the microfluid channel. Gd$^{3+}$ ions in water are slowly pumped into the fluid cell through the tubes. (e) A picture of the microfuild channel on a gold  stripline microwave waveguide. The hBN nanosheet is at the center narrow regime of the gold stripline. The scale bar is 1 cm. (f) ODMR spectra of hBN $V_B^-$ spin defects in water without a magnetic field (blue circles), in water with a 13 mT magnetic field (red squares) and in air without a magnetic field (black triangles). The microwave power is 1 W, and the excitation laser power is 3 mW. The experiment is performed at room temperature.} \label{fig1:schematic}
\end{figure*}

Quantum sensing has emerged as a powerful technique for detecting and measuring a wide range of physical and chemical quantities \cite{degen2017quantum,budker2007optical}. It has been shown to be highly sensitive and have applications in  material science, biology, and medicine \cite{schirhagl2014nitrogen,kucsko2013nanometre,casola2018probing,shi2018single}. Recently, optically active spin defects in hexagonal boron nitride (hBN) \cite{gottscholl2020initialization,gottscholl2021room,mendelson2021identifying,chejanovsky2021single} were emerging as promising platforms for quantum sensing \cite{vaidya2023quantum}. The 2D nature of hBN allows for spin defects to be embedded in atomically thin layers while maintaining high spin qualities\cite{gao2021high}, enabling the sensor to be in close proximity to the target sample, thereby improving sensitivity. Also, a 2D vdW material can be readily integrated into other devices and form multifunctional heterostructures, which opens prospects for in situ quantum sensing \cite{novoselov20162d}. So far, hBN spin defects have been used for sensing multiple physical quantities in solids, including static magnetic fields\cite{huang2022wide,healey2022quantum}, temperature \cite{gottscholl2021spin,liu2021temperature,healey2022quantum},  strain \cite{yang2022spin,lyu2022strain}, and nuclear spins \cite{gao2022nuclear}. However, quantum sensing of paramagnetic ions in liquids with hBN spin defects has not been reported, even though  paramagnetic ions play critical roles in chemical, biological and medical sciences.

Paramagnetic ions and radicals contain at least one unpaired electron, and are involved in various physiological processes including cell signaling \cite{thomas2015breathing} and immune response to infection \cite{bogdan2015nitric}. Some paramagnetic ions can be potentially used as biomarkers for monitoring disease states \cite{griendling2016measurement}. For example, gadolinium ions (Gd$^{3+}$) are widely invoked as relaxation agents and play a major role in magnetic resonance imaging (MRI) \cite{weinmann1984characteristics,chan2007small,doi:10.1148/radiol.2021210957,dos2022thermal}. Iron ions participate in many activities in the human body, including oxygen transport, enzyme function and mitochondrial energy provision \cite{ma2021parkinson,samrot2021review}. The detection of paramagnetic ions under physiological conditions is highly desired due to their multiple roles in biological and medical sciences.

Previous studies have shown optically addressable spin color centers in bulk 3D materials, such nitrogen vacancy (NV) center in diamond, can be used as promising sensors for detecting paramagnetic spins \cite{steinert2013magnetic,ziem2013highly, ermakova2013detection,shi2015single,simpson2017electron,gorrini2019fast,radu2019dynamic}.  The measurements rely on detecting  magnetic noise from fluctuating  spins of paramagnetic ions. However, the noise signal $\Gamma$ decays significantly as the distance $d$ between the sensor and target spins increases, following $\Gamma$ $\propto$ $1/d^6$ for a single target spin. Therefore, these measurements require the sensor to be in close proximity to the target samples.  However, creating high-quality spin color centers near the surface of 3D bulk materials remains challenging due to the inevitable dangling bonds on the surface of bulk 3D materials.

Spin defects in hBN \cite{gottscholl2021spin,liu2022spin} provide a promising solution to the challenges of creating high-quality spin color centers near surfaces \citep{gao2021high,xu2022greatly}. hBN can be stable at the limit of a monolayer and have no dangling bond on the surface. Spin defects can be readily created close to the hBN surface while sustaining high stability and spin properties without any further surface treatment. Moreover, hBN can be exfoliated into thin flakes and produced in large quantities, which can significantly reduce the cost of production. Here we report the first quantum sensing of Gd$^{3+}$ paramagnetic ions in liquids using negatively charged boron vacancy ($V_B^-$) spin defects  in hBN in a microfluid structure (Figure \ref{fig1:schematic}(a)). Employing spin relaxometry, we observe a reduction of $T_1$ relaxation time in the presence of paramagnetic ions. We also present a highly sensitive sensing technique based on the contrast of photoluminescence (PL) emission with and without microwaves. Our results offer new opportunities for the development of highly sensitive and portable sensors for paramagnetic ions with potential applications in medicine, biology, and chemistry.

The $V_B^-$ spin defect is a color center in hBN consisting of a missing boron atom at its lattice site in a negatively charged state \cite{gottscholl2020initialization,ivady2020ab,mathur2022excited,baber2021excited,mu2022excited,yu2022excited} (Figure \ref{fig1:schematic}(b)). The symmetry axis of $V_B^-$ defects is always along the c-axis perpendicular to the hBN 2D lattice, making $V_B^-$ well suited for ensemble measurements.  The $V_B^-$ has a $S$ = 1 triplet ground state with a zero field splitting of $D_{gs}$ $\approx$ 3.47 GHz (Figure \ref{fig1:schematic} (c)). The spin dependent PL emission, together with spin polarization via laser excitation, enables optically detected magnetic resonance (ODMR) experiments. Our sensor consist of an ensemble of $V_B^-$ defects created by low energy (600 eV) helium ions with a large dose density of about $10^{15}$ cm$^{-2}$. The average depth of created $V_B^-$ defects in an hBN nanosheet is 6.4 nm, allowing for a small sensor-sample separation \cite{ziegler2010srim}. The thickness of the hBN nanosheet is typically in the range of 30-50 nm for achieving a strong plasmonic enhancement of the PL intensity on a gold microwave waveguide \citep{gao2021high}.  The lateral size of the hBN nanosheet is usually larger than 10 $\mu$m. 

\begin{figure*}[tbh]
	\centering
	\includegraphics[width=0.85\textwidth]{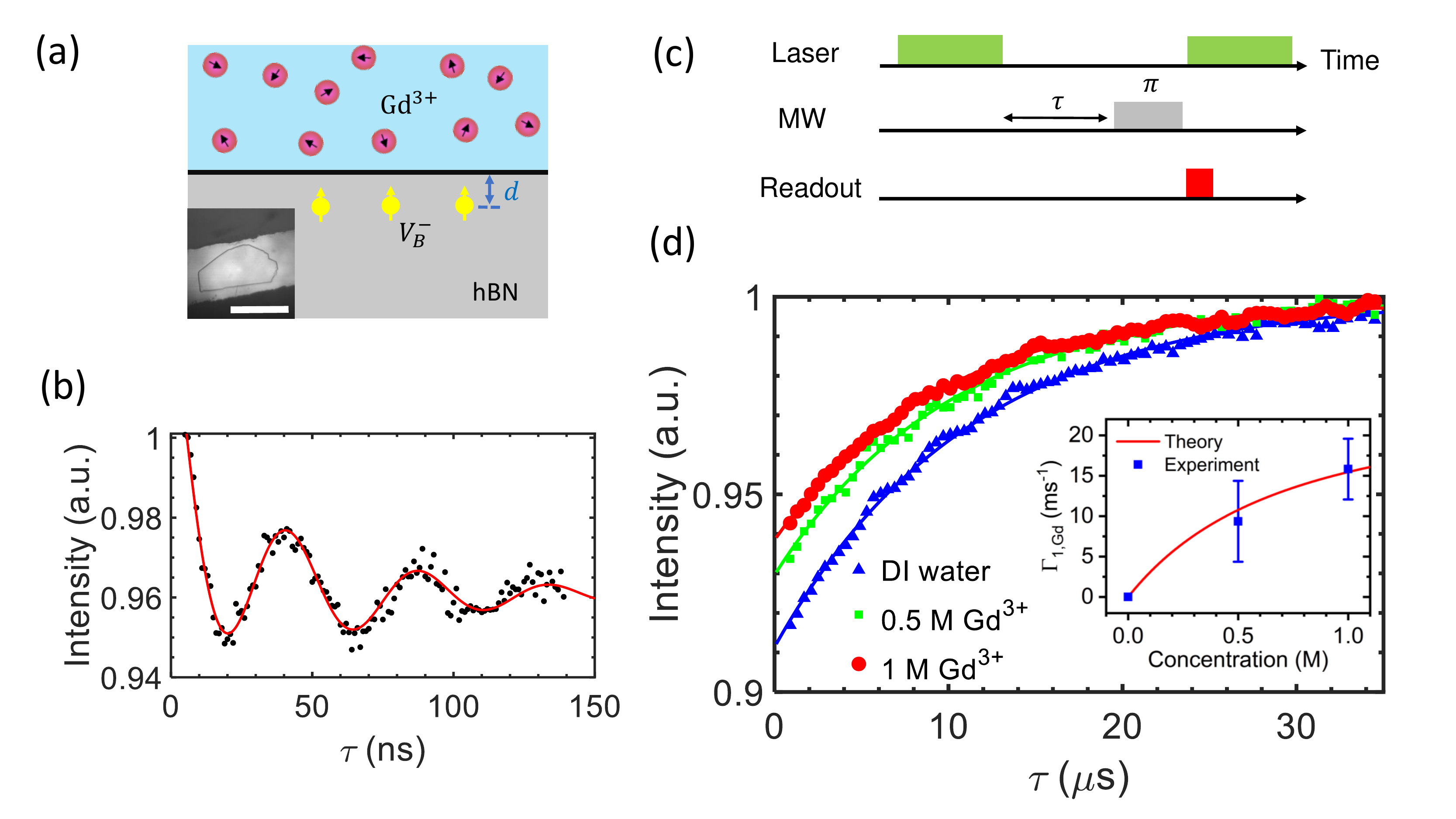}
	\caption{ \textbf{Effects of liquid paramagnetic ions on T1 of shallow hBN spin defects.} (a) An illustration of paramagentic ions in a liquid that induce spin relaxation in nearby spin qubits. The average depth $d$ of $V_B^-$ defects from the top hBN surface is  6.4 nm. Inset: A optical image of an hBN nanosheet on the gold waveguide. The scale bar is 35 $\mu$m. (b) Rabi oscillation of $V_B^-$  spin defects measured in water at 13 mT. The Rabi frequency is fitted as $f_{Rabi}$ = 21.6 MHz. (c) Schematic of the pulse sequence for $T_1$ relaxometry measurements. A initial pumping laser polarizes the $V_B^-$ defect spins to $\ket{m_s=0}$. After a waiting time of $\tau$, a microwave $\pi$ pulse follow by another laser pulse is applied to readout the final spin state. (d) Experimental results of $T_1$ relaxometry in DI water and Gd$^{3+}$ solutions. Two different concentrations of Gd$^{3+}$ are used here. The relaxation time are fitted as 11.24$\pm$0.17 $\mu$s,  10.17$\pm$0.50 $\mu$s, and 9.54$\pm$0.32 $\mu$s for DI water (blue triangles), 0.5 M Gd$^{3+}$ solution (green squares) and 1 M Gd$^{3+}$ solution (red circles), respectively. The inset figure shows experimental and theoretical results of Gd$^{3+}$ induced spin relaxation rates as functions of Gd$^{3+}$  concentration. } \label{fig2:T1}
\end{figure*}

Figure \ref{fig1:schematic}(a)(d) illustrate our experimental setup for sensing paramagnetic ions with shallow spin defects in hBN. First, we transfer an hBN nanosheet with $V_B^-$ defects onto a gold microwave  transmission line with a width of 35 $\mu$m. A coverslip is then placed on top of the microwave waveguide, spaced by double-side tapes. Two pipes are connected to the two sides of the colver slip for delivering and changing solutions, and the entire device is sealed with epoxy to form a micorfluid cell. Figure \ref{fig1:schematic}(e) shows a picture of the prepared device. In the experiment, we use a green laser (532 nm) to excite the $V_B^-$ defects, and collect photon emission with a 750 nm long pass filter. A confocal PL map shows the nearly homogeneous distribution of $V_B^-$ defects over the  hBN nanosheet. We first characterize $V_B^-$ spin defects in the transferred hBN nanosheet in air. We then slowly inject a deionized (DI) water into the microfluid cell. Continuous wave (CW) ODMR measurements confirm that $V_B^-$ defects maintain high-contrast (about 20\%) ODMR signals in the liquid environment (Figure \ref{fig1:schematic}(f)). In the following discussion, all experiments are performed in liquids, and  DI water is used to characterize $V_B^-$ defects as a reference.

We use spin relaxometry to detect surrounding paramagnetic ions by measuring the spin relaxation time ($T_1$) of shallow $V_B^-$ spin defects (Figure \ref{fig2:T1}(a)). In a low magnetic field, free diffusing paramagnetic ions exhibit a zero average magnetic field but non-zero root-mean-square (RMS) field due to spin fluctuations. Such stochastic fields can speed up spin relaxation of  nearby spin qubits. In our experiment, we first perform coherent control of the $V_B^-$ spins and extract the Rabi frequency for generating a $\pi$  pulse for $T_1$ relaxometry measurements. A 1.5-$\mu$s green laser pulse is used for spin initialization and readout, and a microwave pulse is used to drive the spin to oscillate between the $\ket{m_s=0}$ and $\ket{m_s=-1}$ states. As shown in Figure \ref{fig2:T1}(b), by varying the microwave pulse duration, we see a oscillation of the PL intensity with a period of around 46.3 ns ($f_{Rabi}$ = 21.6 MHz). Then the longitudinal relaxation time $T_1$ is characterized by using the pulse sequence as depicted in Figure \ref{fig2:T1}(c). In DI water, the $T_{1,DI}$ is measured to be 11.24$\pm$0.17 $\mu$s, which is close to $T_1$ in air for this hBN nanosheet. They are relatively short due to the high doping density that we used to obtain high PL count rates \cite{gong2022coherent}.  In the presence of Gd$^{3+}$ ions (prepared by dissolving Gd(NO$_3$)$_3$ in water), we observe reduced $T_1$ relaxation times. The measured spin relaxation times are $T_{1}=9.54 \pm 0.32$ $\mu$s for 1~M Gd$^{3+}$ ions and $T_{1}=10.17 \pm 0.50 $ $\mu$s for 0.5 M Gd$^{3+}$ ions (Figure \ref{fig2:T1}(d)). Using DI water as the reference, the Gd$^{3+}$ induced spin relaxation rates can be obtained as $\Gamma_{1,Gd}=1/T_{1}-1/T_{1,DI}$. They are found to be $\Gamma_{1,Gd}=$ 15.8$\pm$3.8 ms$^{-1}$ for 1~M Gd$^{3+}$ ions and $\Gamma_{1,Gd}=$ 9.4$\pm$5.0 ms$^{-1}$ for 0.5 M Gd$^{3+}$ ions. The measured average values are significantly larger than the uncertainties of our measurements, showing that we have detected paramagnetic ions in water with hBN spin defects for the first time.

\begin{figure*}[tbh]
	\centering
	\includegraphics[width=1\textwidth]{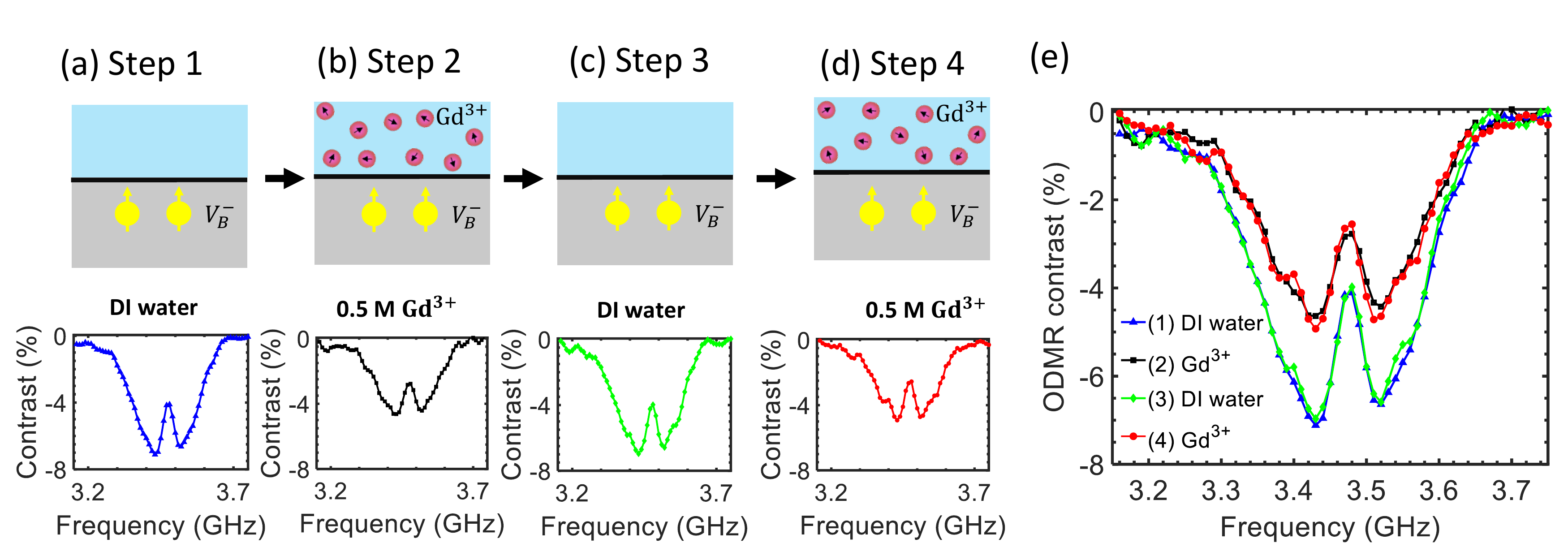}
	\caption{ \textbf{Reproducibility of the reduction of  the ODMR contrast of shallow hBN spin defects due to liquid paramagnetic ions.} Top panels: Illustration of CW ODMR experiments in DI water (a), 0.5 M Gd$^{3+}$ solution (b), DI water (c), and 0.5 M Gd$^{3+}$ solution (d). Step 1-4 are performed in sequence. Bottom panels: CW ODMR results of Step 1-4. Step 2 and Step 4 show the same amount of reduction in ODMR contrasts due to 0.5 M Gd$^{3+}$ solutions. (e) A summary of the ODMR spectra in (a)-(d) for comparison. No external magnetic field is applied. } \label{fig3:reprodcibility}
\end{figure*} 	 

The paramagnetic-ion-induced spin relaxation rates $\Gamma_{1,Gd}$ depends on the corresponding RMS magnetic field and its spectral density $S_{Gd}(\omega)=\sqrt{2/\pi}\cdot \omega_{Gd}/[(\omega-\omega_L)^2+\omega_{Gd}^2]$, where $\omega_L$ is the Larmor frequency of Gd$^{3+}$. $\omega_{Gd}$ is the composite
relaxation rate of Gd$^{3+}$ spins, which is on the order of tens of gigahertz. In a low magnetic field, $\omega_L$ is negligible compares to $\omega_{Gd}$, and hence $S_{Gd}$ is dominated by statistical polarization and substantial broadening effects of  fluctuations \cite{steinert2013magnetic}. When the thickness of the Gd$^{3+}$ solution is far larger than the average depth $d$ of $V_B^-$ spins, we can assume that Gd$^{3+}$ ions exist everywhere in the half infinite space above hBN. Then the Gd$^{3+}$ induced decay rate $\Gamma_{1,Gd}$ of $V_B^-$ spins is \cite{steinert2013magnetic}
\begin{equation}
	\Gamma_{1,Gd} \approx \frac{21\cdot 10^3\pi N_A C_{Gd}}{8d^3}\frac{(\pi\mu_0 \hbar\gamma_{e}\gamma_{Gd})^2 \omega_{Gd}}{\omega_{Gd}^2+4\pi^2 D_{gs}^2},
	\label{Eq1:relaxation}
\end{equation}
where $C_{Gd}$ is the Gd$^{3+}$ concentration in mol$\cdot$l$^{-1}$, $N_A$ is the Avogadro number, $\mu_0$ is the vacuum permeability, $d$ is average mean depth of the $V_B^-$ defects, and $\gamma_{Gd}\approx\gamma_{e}\approx28.0$ GHz/T is the gyromagnetic ratio of electrons. $\omega_{Gd} \approx 50 \times 10^9$ s$^{-1}$ + $C_{Gd} \cdot$ (77$\times 10^9$ s$^{-1}$M$^{-1}$) \cite{steinert2013magnetic}. Invoking this relaxation model, the theoretical prediction shows good agreement with the experimental results as shown in the inset of Figure \ref{fig2:T1}(d). This agreement further confirms that we have observed paramagnetic spin noise in liquids with hBN spin defects.
 	 
\begin{figure}[b]
	\centering
	\includegraphics[width=0.5\textwidth]{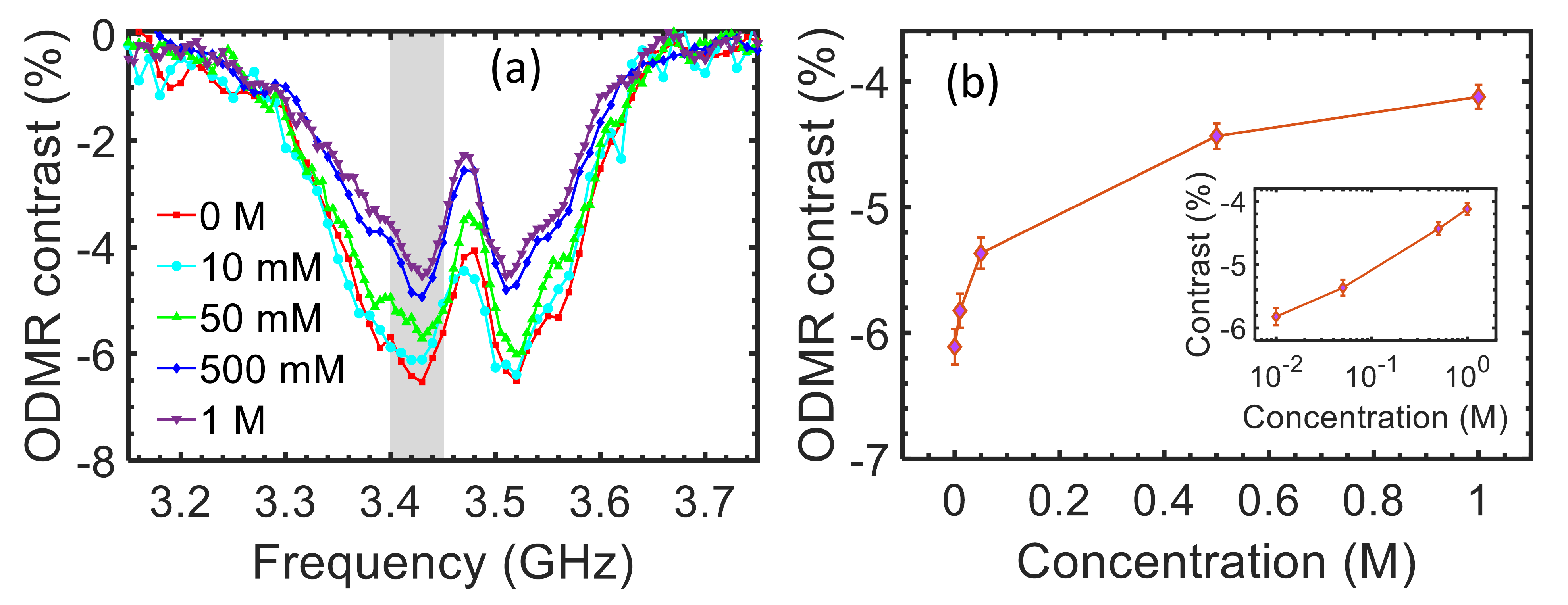}
	\caption{ \textbf{Effects of Gd$^{3+}$ concentration on the $V_B^-$ ODMR contrast.}(a) ODMR spectra for several different Gd$^{3+}$ ion concentrations. All measurements are performed under a 50 mW microwave drive. (b) The averaged ODMR contrast in the  3.40 GHz to 3.45 GHz frequency range (gray box in (a)) as a function of Gd$^{3+}$ ion concentration. } \label{fig4:concentration}
\end{figure}
 	 
After demonstrating the capability of spin relaxometry, we develop another detection method based on the reduction of the ODMR contrast of hBN spin defects due to paramagnetic ions. In the pulsed spin relaxometry measurements, we observe a decrease in the contrast of $V_B^-$ spin relaxation signal in the presence of Gd$^{3+}$ ions  (Figure \ref{fig2:T1}(d)). Similar phenomena can also be observed in CW ODMR experiments. In DI water, $V_B^-$ defects exhibit $\sim$ 7$\%$ ODMR contrast under a 50 mW microwave driving (Figure \ref{fig3:reprodcibility}(a)). Once the 0.5 M Gd$^{3+}$ water solution replaces the DI water, we observe a dramatic reduction of ODMR contrast to less than 5$\%$   (Figure \ref{fig3:reprodcibility}(b)). The ODMR contrast will resume after pumping the DI water back into the microfluid cell (Figure \ref{fig3:reprodcibility}(c)) and will decrease again in the presence of Gd$^{3+}$ solution (Figure \ref{fig3:reprodcibility}(d)). Thus the reduction of ODMR contrast due to paramagnetic ions is reversible and repeatable (Figure \ref{fig3:reprodcibility}(e)). This method allows us to detect paramagnetic ions by monitoring the change in CW ODMR contrasts.

To quantify the dependence of the ODMR contrast on paramagnetic ion concentration, we perform the CW ODMR experiments in five different solutions: 0 mM Gd$^{3+}$ (DI water), 10 mM Gd$^{3+}$, 50 mM Gd$^{3+}$, 500 mM Gd$^{3+}$, and 1 M Gd$^{3+}$. As the concentration of Gd$^{3+}$ increases, we observe a continuous reduction in the ODMR contrast (Figure \ref{fig4:concentration}). To quantify this concentration dependency, we select data points within the frequency range from 3.40 GHz to 3.45 GHz and calculate the average ODMR contrasts (Figure \ref{fig4:concentration}(b)). The effect of Gd$^{3+}$ ions at a concentration as low as 10 mM is observable.  
This method requires only a few ODMR data points to calculate the average integrated ODMR contrast and does not rely on pulsed laser excitation, precise microwave manipulation, fast photodetectors, or multichannel pulse generators. This simplicity allows for rapid detection of paramagnetic ions in liquids, making it ideal for real-world applications.

The reduction of the ODMR contrast can be explained by the Gd$^{3+}$ induced depopulation of the $m_s=0$ ground state during the laser initialization, which has also been observed with diamond NV centers \cite{gorrini2019fast,radu2019dynamic}. While a green laser tries to initialize  spin defects to the $m_s=0$ state, high-frequency magnetic noise will drive spin transitions from the $m_s=0$ state to the $m_s=\pm 1$ states. In a  solution of paramagnetic ions, a higher spin concentration gives rise to more magnetic noise due to spin fluctuations. A stronger spin relaxation effect due to spin noise will result in a lower spin polarization level for a given laser power, which gives rise to the concentration-dependent  ODMR contrast reduction in shallow hBN spin defects. Additionally, Gd$^{3+}$ ions can affect solution conductivity and cause slight microwave absorption, which can also impact the ODMR contrast of hBN spin defects.  In the future, to mitigate the potential microwave absorption effect of Gd$^{3+}$ ions, we can reduce the width of the microfluid channel to expose only a small portion of the microwave waveguide to the ionic solution.  Furthermore, we can employ two hBN nanosheets with spin defects at two different depths: shallow spin defects can detect the spin noise of paramagnetic ions, while deep spin defects can monitor any potential ODMR contrast reduction due to microwave absorption by the ionic solution. Overall, the concentration-sensitive ODMR contrast allows for efficient detection of paramagnetic ions in liquids using CW ODMR, which is much simpler to implement than pulsed sensing protocols.

In conclusion, we have demonstrated the first detection of spin noise from paramagnetic ions in liquids using a vdW sensor based on hBN spin defects. Our spin relaxometry measurements reveal the characteristic behavior of the Gd$^{3+}$-induced spin relaxation, which increases with increasing Gd$^{3+}$ concentration.
 By using CW ODMR technique, we are also able to detect the Gd$^{3+}$ ions via the contrast reduction of the spin-dependent PL. 
Our work represents an initial demonstration of the potential of hBN spin defects for paramagnetic ion sensing in liquids. There are many ways for further improvement. For instance, the sensitivity can be further improved by optimizing the hBN sensor, including reducing the depth of spin defects and improving the collection efficiency of the spin-dependent fluorescence. Future studies can also explore the sensing performance of hBN spin defects in more complex chemical and biological  environments.

Note added: While finalizing this paper, we became aware of a related paper on detecting paramagnetic ions with hBN spin defects \cite{robertson2023detection}

\section* {Acknowledgments}
T.L. thanks the Purdue Quantum Science and Engineering  Institute (PQSEI) for support through the seed grant, and the DARPA ARRIVE program. Y.P.C. acknowledges support  by the Quantum Science Center, a U.S. Department of Energy, Office of Science, National Quantum Information Science Research Center. 

%\vspace{1cm}
%\bibliographystyle{unsrtnat}
%\bibliography{hBNquantumsensing}

\begin{thebibliography}{46}%
\makeatletter
\providecommand \@ifxundefined [1]{%
 \@ifx{#1\undefined}
}%
\providecommand \@ifnum [1]{%
 \ifnum #1\expandafter \@firstoftwo
 \else \expandafter \@secondoftwo
 \fi
}%
\providecommand \@ifx [1]{%
 \ifx #1\expandafter \@firstoftwo
 \else \expandafter \@secondoftwo
 \fi
}%
\providecommand \natexlab [1]{#1}%
\providecommand \enquote  [1]{``#1''}%
\providecommand \bibnamefont  [1]{#1}%
\providecommand \bibfnamefont [1]{#1}%
\providecommand \citenamefont [1]{#1}%
\providecommand \href@noop [0]{\@secondoftwo}%
\providecommand \href [0]{\begingroup \@sanitize@url \@href}%
\providecommand \@href[1]{\@@startlink{#1}\@@href}%
\providecommand \@@href[1]{\endgroup#1\@@endlink}%
\providecommand \@sanitize@url [0]{\catcode `\\12\catcode `\$12\catcode
  `\&12\catcode `\#12\catcode `\^12\catcode `\_12\catcode `\%12\relax}%
\providecommand \@@startlink[1]{}%
\providecommand \@@endlink[0]{}%
\providecommand \url  [0]{\begingroup\@sanitize@url \@url }%
\providecommand \@url [1]{\endgroup\@href {#1}{\urlprefix }}%
\providecommand \urlprefix  [0]{URL }%
\providecommand \Eprint [0]{\href }%
\providecommand \doibase [0]{https://doi.org/}%
\providecommand \selectlanguage [0]{\@gobble}%
\providecommand \bibinfo  [0]{\@secondoftwo}%
\providecommand \bibfield  [0]{\@secondoftwo}%
\providecommand \translation [1]{[#1]}%
\providecommand \BibitemOpen [0]{}%
\providecommand \bibitemStop [0]{}%
\providecommand \bibitemNoStop [0]{.\EOS\space}%
\providecommand \EOS [0]{\spacefactor3000\relax}%
\providecommand \BibitemShut  [1]{\csname bibitem#1\endcsname}%
\let\auto@bib@innerbib\@empty
%</preamble>
\bibitem [{\citenamefont {Degen}\ \emph {et~al.}(2017)\citenamefont {Degen},
  \citenamefont {Reinhard},\ and\ \citenamefont
  {Cappellaro}}]{degen2017quantum}%
  \BibitemOpen
  \bibfield  {author} {\bibinfo {author} {\bibfnamefont {C.~L.}\ \bibnamefont
  {Degen}}, \bibinfo {author} {\bibfnamefont {F.}~\bibnamefont {Reinhard}},\
  and\ \bibinfo {author} {\bibfnamefont {P.}~\bibnamefont {Cappellaro}},\
  }\bibfield  {title} {\bibinfo {title} {Quantum sensing},\ }\href@noop {}
  {\bibfield  {journal} {\bibinfo  {journal} {Reviews of modern physics}\
  }\textbf {\bibinfo {volume} {89}},\ \bibinfo {pages} {035002} (\bibinfo
  {year} {2017})}\BibitemShut {NoStop}%
\bibitem [{\citenamefont {Budker}\ and\ \citenamefont
  {Romalis}(2007)}]{budker2007optical}%
  \BibitemOpen
  \bibfield  {author} {\bibinfo {author} {\bibfnamefont {D.}~\bibnamefont
  {Budker}}\ and\ \bibinfo {author} {\bibfnamefont {M.}~\bibnamefont
  {Romalis}},\ }\bibfield  {title} {\bibinfo {title} {Optical magnetometry},\
  }\href@noop {} {\bibfield  {journal} {\bibinfo  {journal} {Nature physics}\
  }\textbf {\bibinfo {volume} {3}},\ \bibinfo {pages} {227} (\bibinfo {year}
  {2007})}\BibitemShut {NoStop}%
\bibitem [{\citenamefont {Schirhagl}\ \emph {et~al.}(2014)\citenamefont
  {Schirhagl}, \citenamefont {Chang}, \citenamefont {Loretz},\ and\
  \citenamefont {Degen}}]{schirhagl2014nitrogen}%
  \BibitemOpen
  \bibfield  {author} {\bibinfo {author} {\bibfnamefont {R.}~\bibnamefont
  {Schirhagl}}, \bibinfo {author} {\bibfnamefont {K.}~\bibnamefont {Chang}},
  \bibinfo {author} {\bibfnamefont {M.}~\bibnamefont {Loretz}},\ and\ \bibinfo
  {author} {\bibfnamefont {C.~L.}\ \bibnamefont {Degen}},\ }\bibfield  {title}
  {\bibinfo {title} {Nitrogen-vacancy centers in diamond: nanoscale sensors for
  physics and biology},\ }\href@noop {} {\bibfield  {journal} {\bibinfo
  {journal} {Annu. Rev. Phys. Chem}\ }\textbf {\bibinfo {volume} {65}},\
  \bibinfo {pages} {83} (\bibinfo {year} {2014})}\BibitemShut {NoStop}%
\bibitem [{\citenamefont {Kucsko}\ \emph {et~al.}(2013)\citenamefont {Kucsko},
  \citenamefont {Maurer}, \citenamefont {Yao}, \citenamefont {Kubo},
  \citenamefont {Noh}, \citenamefont {Lo}, \citenamefont {Park},\ and\
  \citenamefont {Lukin}}]{kucsko2013nanometre}%
  \BibitemOpen
  \bibfield  {author} {\bibinfo {author} {\bibfnamefont {G.}~\bibnamefont
  {Kucsko}}, \bibinfo {author} {\bibfnamefont {P.~C.}\ \bibnamefont {Maurer}},
  \bibinfo {author} {\bibfnamefont {N.~Y.}\ \bibnamefont {Yao}}, \bibinfo
  {author} {\bibfnamefont {M.}~\bibnamefont {Kubo}}, \bibinfo {author}
  {\bibfnamefont {H.~J.}\ \bibnamefont {Noh}}, \bibinfo {author} {\bibfnamefont
  {P.~K.}\ \bibnamefont {Lo}}, \bibinfo {author} {\bibfnamefont
  {H.}~\bibnamefont {Park}},\ and\ \bibinfo {author} {\bibfnamefont {M.~D.}\
  \bibnamefont {Lukin}},\ }\bibfield  {title} {\bibinfo {title}
  {Nanometre-scale thermometry in a living cell},\ }\href@noop {} {\bibfield
  {journal} {\bibinfo  {journal} {Nature}\ }\textbf {\bibinfo {volume} {500}},\
  \bibinfo {pages} {54} (\bibinfo {year} {2013})}\BibitemShut {NoStop}%
\bibitem [{\citenamefont {Casola}\ \emph {et~al.}(2018)\citenamefont {Casola},
  \citenamefont {Van Der~Sar},\ and\ \citenamefont
  {Yacoby}}]{casola2018probing}%
  \BibitemOpen
  \bibfield  {author} {\bibinfo {author} {\bibfnamefont {F.}~\bibnamefont
  {Casola}}, \bibinfo {author} {\bibfnamefont {T.}~\bibnamefont {Van
  Der~Sar}},\ and\ \bibinfo {author} {\bibfnamefont {A.}~\bibnamefont
  {Yacoby}},\ }\bibfield  {title} {\bibinfo {title} {Probing condensed matter
  physics with magnetometry based on nitrogen-vacancy centres in diamond},\
  }\href@noop {} {\bibfield  {journal} {\bibinfo  {journal} {Nature Reviews
  Materials}\ }\textbf {\bibinfo {volume} {3}},\ \bibinfo {pages} {17088}
  (\bibinfo {year} {2018})}\BibitemShut {NoStop}%
\bibitem [{\citenamefont {Shi}\ \emph {et~al.}(2018)\citenamefont {Shi},
  \citenamefont {Kong}, \citenamefont {Zhao}, \citenamefont {Zhang},
  \citenamefont {Chen}, \citenamefont {Chen}, \citenamefont {Zhang},
  \citenamefont {Wang}, \citenamefont {Ye}, \citenamefont {Wang} \emph
  {et~al.}}]{shi2018single}%
  \BibitemOpen
  \bibfield  {author} {\bibinfo {author} {\bibfnamefont {F.}~\bibnamefont
  {Shi}}, \bibinfo {author} {\bibfnamefont {F.}~\bibnamefont {Kong}}, \bibinfo
  {author} {\bibfnamefont {P.}~\bibnamefont {Zhao}}, \bibinfo {author}
  {\bibfnamefont {X.}~\bibnamefont {Zhang}}, \bibinfo {author} {\bibfnamefont
  {M.}~\bibnamefont {Chen}}, \bibinfo {author} {\bibfnamefont {S.}~\bibnamefont
  {Chen}}, \bibinfo {author} {\bibfnamefont {Q.}~\bibnamefont {Zhang}},
  \bibinfo {author} {\bibfnamefont {M.}~\bibnamefont {Wang}}, \bibinfo {author}
  {\bibfnamefont {X.}~\bibnamefont {Ye}}, \bibinfo {author} {\bibfnamefont
  {Z.}~\bibnamefont {Wang}}, \emph {et~al.},\ }\bibfield  {title} {\bibinfo
  {title} {Single-dna electron spin resonance spectroscopy in aqueous
  solutions},\ }\href@noop {} {\bibfield  {journal} {\bibinfo  {journal}
  {Nature methods}\ }\textbf {\bibinfo {volume} {15}},\ \bibinfo {pages} {697}
  (\bibinfo {year} {2018})}\BibitemShut {NoStop}%
\bibitem [{\citenamefont {Gottscholl}\ \emph {et~al.}(2020)\citenamefont
  {Gottscholl}, \citenamefont {Kianinia}, \citenamefont {Soltamov},
  \citenamefont {Orlinskii}, \citenamefont {Mamin}, \citenamefont {Bradac},
  \citenamefont {Kasper}, \citenamefont {Krambrock}, \citenamefont {Sperlich},
  \citenamefont {Toth} \emph {et~al.}}]{gottscholl2020initialization}%
  \BibitemOpen
  \bibfield  {author} {\bibinfo {author} {\bibfnamefont {A.}~\bibnamefont
  {Gottscholl}}, \bibinfo {author} {\bibfnamefont {M.}~\bibnamefont
  {Kianinia}}, \bibinfo {author} {\bibfnamefont {V.}~\bibnamefont {Soltamov}},
  \bibinfo {author} {\bibfnamefont {S.}~\bibnamefont {Orlinskii}}, \bibinfo
  {author} {\bibfnamefont {G.}~\bibnamefont {Mamin}}, \bibinfo {author}
  {\bibfnamefont {C.}~\bibnamefont {Bradac}}, \bibinfo {author} {\bibfnamefont
  {C.}~\bibnamefont {Kasper}}, \bibinfo {author} {\bibfnamefont
  {K.}~\bibnamefont {Krambrock}}, \bibinfo {author} {\bibfnamefont
  {A.}~\bibnamefont {Sperlich}}, \bibinfo {author} {\bibfnamefont
  {M.}~\bibnamefont {Toth}}, \emph {et~al.},\ }\bibfield  {title} {\bibinfo
  {title} {Initialization and read-out of intrinsic spin defects in a van der
  {Waals} crystal at room temperature},\ }\href@noop {} {\bibfield  {journal}
  {\bibinfo  {journal} {Nature materials}\ }\textbf {\bibinfo {volume} {19}},\
  \bibinfo {pages} {540} (\bibinfo {year} {2020})}\BibitemShut {NoStop}%
\bibitem [{\citenamefont {Gottscholl}\ \emph
  {et~al.}(2021{\natexlab{a}})\citenamefont {Gottscholl}, \citenamefont {Diez},
  \citenamefont {Soltamov}, \citenamefont {Kasper}, \citenamefont {Sperlich},
  \citenamefont {Kianinia}, \citenamefont {Bradac}, \citenamefont
  {Aharonovich},\ and\ \citenamefont {Dyakonov}}]{gottscholl2021room}%
  \BibitemOpen
  \bibfield  {author} {\bibinfo {author} {\bibfnamefont {A.}~\bibnamefont
  {Gottscholl}}, \bibinfo {author} {\bibfnamefont {M.}~\bibnamefont {Diez}},
  \bibinfo {author} {\bibfnamefont {V.}~\bibnamefont {Soltamov}}, \bibinfo
  {author} {\bibfnamefont {C.}~\bibnamefont {Kasper}}, \bibinfo {author}
  {\bibfnamefont {A.}~\bibnamefont {Sperlich}}, \bibinfo {author}
  {\bibfnamefont {M.}~\bibnamefont {Kianinia}}, \bibinfo {author}
  {\bibfnamefont {C.}~\bibnamefont {Bradac}}, \bibinfo {author} {\bibfnamefont
  {I.}~\bibnamefont {Aharonovich}},\ and\ \bibinfo {author} {\bibfnamefont
  {V.}~\bibnamefont {Dyakonov}},\ }\bibfield  {title} {\bibinfo {title} {Room
  temperature coherent control of spin defects in hexagonal boron nitride},\
  }\href@noop {} {\bibfield  {journal} {\bibinfo  {journal} {Science Advances}\
  }\textbf {\bibinfo {volume} {7}},\ \bibinfo {pages} {eabf3630} (\bibinfo
  {year} {2021}{\natexlab{a}})}\BibitemShut {NoStop}%
\bibitem [{\citenamefont {Mendelson}\ \emph {et~al.}(2021)\citenamefont
  {Mendelson}, \citenamefont {Chugh}, \citenamefont {Reimers}, \citenamefont
  {Cheng}, \citenamefont {Gottscholl}, \citenamefont {Long}, \citenamefont
  {Mellor}, \citenamefont {Zettl}, \citenamefont {Dyakonov}, \citenamefont
  {Beton} \emph {et~al.}}]{mendelson2021identifying}%
  \BibitemOpen
  \bibfield  {author} {\bibinfo {author} {\bibfnamefont {N.}~\bibnamefont
  {Mendelson}}, \bibinfo {author} {\bibfnamefont {D.}~\bibnamefont {Chugh}},
  \bibinfo {author} {\bibfnamefont {J.~R.}\ \bibnamefont {Reimers}}, \bibinfo
  {author} {\bibfnamefont {T.~S.}\ \bibnamefont {Cheng}}, \bibinfo {author}
  {\bibfnamefont {A.}~\bibnamefont {Gottscholl}}, \bibinfo {author}
  {\bibfnamefont {H.}~\bibnamefont {Long}}, \bibinfo {author} {\bibfnamefont
  {C.~J.}\ \bibnamefont {Mellor}}, \bibinfo {author} {\bibfnamefont
  {A.}~\bibnamefont {Zettl}}, \bibinfo {author} {\bibfnamefont
  {V.}~\bibnamefont {Dyakonov}}, \bibinfo {author} {\bibfnamefont {P.~H.}\
  \bibnamefont {Beton}}, \emph {et~al.},\ }\bibfield  {title} {\bibinfo {title}
  {Identifying carbon as the source of visible single-photon emission from
  hexagonal boron nitride},\ }\href@noop {} {\bibfield  {journal} {\bibinfo
  {journal} {Nature materials}\ }\textbf {\bibinfo {volume} {20}},\ \bibinfo
  {pages} {321} (\bibinfo {year} {2021})}\BibitemShut {NoStop}%
\bibitem [{\citenamefont {Chejanovsky}\ \emph {et~al.}(2021)\citenamefont
  {Chejanovsky}, \citenamefont {Mukherjee}, \citenamefont {Geng}, \citenamefont
  {Chen}, \citenamefont {Kim}, \citenamefont {Denisenko}, \citenamefont
  {Finkler}, \citenamefont {Taniguchi}, \citenamefont {Watanabe}, \citenamefont
  {Dasari} \emph {et~al.}}]{chejanovsky2021single}%
  \BibitemOpen
  \bibfield  {author} {\bibinfo {author} {\bibfnamefont {N.}~\bibnamefont
  {Chejanovsky}}, \bibinfo {author} {\bibfnamefont {A.}~\bibnamefont
  {Mukherjee}}, \bibinfo {author} {\bibfnamefont {J.}~\bibnamefont {Geng}},
  \bibinfo {author} {\bibfnamefont {Y.-C.}\ \bibnamefont {Chen}}, \bibinfo
  {author} {\bibfnamefont {Y.}~\bibnamefont {Kim}}, \bibinfo {author}
  {\bibfnamefont {A.}~\bibnamefont {Denisenko}}, \bibinfo {author}
  {\bibfnamefont {A.}~\bibnamefont {Finkler}}, \bibinfo {author} {\bibfnamefont
  {T.}~\bibnamefont {Taniguchi}}, \bibinfo {author} {\bibfnamefont
  {K.}~\bibnamefont {Watanabe}}, \bibinfo {author} {\bibfnamefont {D.~B.~R.}\
  \bibnamefont {Dasari}}, \emph {et~al.},\ }\bibfield  {title} {\bibinfo
  {title} {Single-spin resonance in a van der {Waals} embedded paramagnetic
  defect},\ }\href@noop {} {\bibfield  {journal} {\bibinfo  {journal} {Nature
  materials}\ }\textbf {\bibinfo {volume} {20}},\ \bibinfo {pages} {1079}
  (\bibinfo {year} {2021})}\BibitemShut {NoStop}%
\bibitem [{\citenamefont {Vaidya}\ \emph {et~al.}(2023)\citenamefont {Vaidya},
  \citenamefont {Gao}, \citenamefont {Dikshit}, \citenamefont {Aharonovich},\
  and\ \citenamefont {Li}}]{vaidya2023quantum}%
  \BibitemOpen
  \bibfield  {author} {\bibinfo {author} {\bibfnamefont {S.}~\bibnamefont
  {Vaidya}}, \bibinfo {author} {\bibfnamefont {X.}~\bibnamefont {Gao}},
  \bibinfo {author} {\bibfnamefont {S.}~\bibnamefont {Dikshit}}, \bibinfo
  {author} {\bibfnamefont {I.}~\bibnamefont {Aharonovich}},\ and\ \bibinfo
  {author} {\bibfnamefont {T.}~\bibnamefont {Li}},\ }\bibfield  {title}
  {\bibinfo {title} {Quantum sensing and imaging with spin defects in hexagonal
  boron nitride},\ }\href@noop {} {\bibfield  {journal} {\bibinfo  {journal}
  {arXiv preprint arXiv:2302.11169}\ } (\bibinfo {year} {2023})}\BibitemShut
  {NoStop}%
\bibitem [{\citenamefont {Gao}\ \emph {et~al.}(2021)\citenamefont {Gao},
  \citenamefont {Jiang}, \citenamefont {Llacsahuanga~Allcca}, \citenamefont
  {Shen}, \citenamefont {Sadi}, \citenamefont {Solanki}, \citenamefont {Ju},
  \citenamefont {Xu}, \citenamefont {Upadhyaya}, \citenamefont {Chen} \emph
  {et~al.}}]{gao2021high}%
  \BibitemOpen
  \bibfield  {author} {\bibinfo {author} {\bibfnamefont {X.}~\bibnamefont
  {Gao}}, \bibinfo {author} {\bibfnamefont {B.}~\bibnamefont {Jiang}}, \bibinfo
  {author} {\bibfnamefont {A.~E.}\ \bibnamefont {Llacsahuanga~Allcca}},
  \bibinfo {author} {\bibfnamefont {K.}~\bibnamefont {Shen}}, \bibinfo {author}
  {\bibfnamefont {M.~A.}\ \bibnamefont {Sadi}}, \bibinfo {author}
  {\bibfnamefont {A.~B.}\ \bibnamefont {Solanki}}, \bibinfo {author}
  {\bibfnamefont {P.}~\bibnamefont {Ju}}, \bibinfo {author} {\bibfnamefont
  {Z.}~\bibnamefont {Xu}}, \bibinfo {author} {\bibfnamefont {P.}~\bibnamefont
  {Upadhyaya}}, \bibinfo {author} {\bibfnamefont {Y.~P.}\ \bibnamefont {Chen}},
  \emph {et~al.},\ }\bibfield  {title} {\bibinfo {title} {High-contrast
  plasmonic-enhanced shallow spin defects in hexagonal boron nitride for
  quantum sensing},\ }\href@noop {} {\bibfield  {journal} {\bibinfo  {journal}
  {Nano Letters}\ }\textbf {\bibinfo {volume} {21}},\ \bibinfo {pages} {7708}
  (\bibinfo {year} {2021})}\BibitemShut {NoStop}%
\bibitem [{\citenamefont {Novoselov}\ \emph {et~al.}(2016)\citenamefont
  {Novoselov}, \citenamefont {Mishchenko}, \citenamefont {Carvalho},\ and\
  \citenamefont {Castro~Neto}}]{novoselov20162d}%
  \BibitemOpen
  \bibfield  {author} {\bibinfo {author} {\bibfnamefont {K.~S.}\ \bibnamefont
  {Novoselov}}, \bibinfo {author} {\bibfnamefont {A.}~\bibnamefont
  {Mishchenko}}, \bibinfo {author} {\bibfnamefont {A.}~\bibnamefont
  {Carvalho}},\ and\ \bibinfo {author} {\bibfnamefont {A.~H.}\ \bibnamefont
  {Castro~Neto}},\ }\bibfield  {title} {\bibinfo {title} {{2D} materials and
  van der {Waals} heterostructures},\ }\href@noop {} {\bibfield  {journal}
  {\bibinfo  {journal} {Science}\ }\textbf {\bibinfo {volume} {353}},\ \bibinfo
  {pages} {aac9439} (\bibinfo {year} {2016})}\BibitemShut {NoStop}%
\bibitem [{\citenamefont {Huang}\ \emph {et~al.}(2022)\citenamefont {Huang},
  \citenamefont {Zhou}, \citenamefont {Chen}, \citenamefont {Lu}, \citenamefont
  {McLaughlin}, \citenamefont {Li}, \citenamefont {Alghamdi}, \citenamefont
  {Djugba}, \citenamefont {Shi}, \citenamefont {Wang} \emph
  {et~al.}}]{huang2022wide}%
  \BibitemOpen
  \bibfield  {author} {\bibinfo {author} {\bibfnamefont {M.}~\bibnamefont
  {Huang}}, \bibinfo {author} {\bibfnamefont {J.}~\bibnamefont {Zhou}},
  \bibinfo {author} {\bibfnamefont {D.}~\bibnamefont {Chen}}, \bibinfo {author}
  {\bibfnamefont {H.}~\bibnamefont {Lu}}, \bibinfo {author} {\bibfnamefont
  {N.~J.}\ \bibnamefont {McLaughlin}}, \bibinfo {author} {\bibfnamefont
  {S.}~\bibnamefont {Li}}, \bibinfo {author} {\bibfnamefont {M.}~\bibnamefont
  {Alghamdi}}, \bibinfo {author} {\bibfnamefont {D.}~\bibnamefont {Djugba}},
  \bibinfo {author} {\bibfnamefont {J.}~\bibnamefont {Shi}}, \bibinfo {author}
  {\bibfnamefont {H.}~\bibnamefont {Wang}}, \emph {et~al.},\ }\bibfield
  {title} {\bibinfo {title} {Wide field imaging of van der {Waals} ferromagnet
  {Fe$_3$GeTe$_2$} by spin defects in hexagonal boron nitride},\ }\href@noop {}
  {\bibfield  {journal} {\bibinfo  {journal} {Nature communications}\ }\textbf
  {\bibinfo {volume} {13}},\ \bibinfo {pages} {5369} (\bibinfo {year}
  {2022})}\BibitemShut {NoStop}%
\bibitem [{\citenamefont {Healey}\ \emph {et~al.}(2023)\citenamefont {Healey},
  \citenamefont {Scholten}, \citenamefont {Yang}, \citenamefont {Scott},
  \citenamefont {Abrahams}, \citenamefont {Robertson}, \citenamefont {Hou},
  \citenamefont {Guo}, \citenamefont {Rahman}, \citenamefont {Lu} \emph
  {et~al.}}]{healey2022quantum}%
  \BibitemOpen
  \bibfield  {author} {\bibinfo {author} {\bibfnamefont {A.}~\bibnamefont
  {Healey}}, \bibinfo {author} {\bibfnamefont {S.}~\bibnamefont {Scholten}},
  \bibinfo {author} {\bibfnamefont {T.}~\bibnamefont {Yang}}, \bibinfo {author}
  {\bibfnamefont {J.}~\bibnamefont {Scott}}, \bibinfo {author} {\bibfnamefont
  {G.}~\bibnamefont {Abrahams}}, \bibinfo {author} {\bibfnamefont
  {I.}~\bibnamefont {Robertson}}, \bibinfo {author} {\bibfnamefont
  {X.}~\bibnamefont {Hou}}, \bibinfo {author} {\bibfnamefont {Y.}~\bibnamefont
  {Guo}}, \bibinfo {author} {\bibfnamefont {S.}~\bibnamefont {Rahman}},
  \bibinfo {author} {\bibfnamefont {Y.}~\bibnamefont {Lu}}, \emph {et~al.},\
  }\bibfield  {title} {\bibinfo {title} {Quantum microscopy with van der
  {Waals} heterostructures},\ }\href@noop {} {\bibfield  {journal} {\bibinfo
  {journal} {Nature Physics}\ }\textbf {\bibinfo {volume} {19}},\ \bibinfo
  {pages} {87–91} (\bibinfo {year} {2023})}\BibitemShut {NoStop}%
\bibitem [{\citenamefont {Gottscholl}\ \emph
  {et~al.}(2021{\natexlab{b}})\citenamefont {Gottscholl}, \citenamefont {Diez},
  \citenamefont {Soltamov}, \citenamefont {Kasper}, \citenamefont {Krau{\ss}e},
  \citenamefont {Sperlich}, \citenamefont {Kianinia}, \citenamefont {Bradac},
  \citenamefont {Aharonovich},\ and\ \citenamefont
  {Dyakonov}}]{gottscholl2021spin}%
  \BibitemOpen
  \bibfield  {author} {\bibinfo {author} {\bibfnamefont {A.}~\bibnamefont
  {Gottscholl}}, \bibinfo {author} {\bibfnamefont {M.}~\bibnamefont {Diez}},
  \bibinfo {author} {\bibfnamefont {V.}~\bibnamefont {Soltamov}}, \bibinfo
  {author} {\bibfnamefont {C.}~\bibnamefont {Kasper}}, \bibinfo {author}
  {\bibfnamefont {D.}~\bibnamefont {Krau{\ss}e}}, \bibinfo {author}
  {\bibfnamefont {A.}~\bibnamefont {Sperlich}}, \bibinfo {author}
  {\bibfnamefont {M.}~\bibnamefont {Kianinia}}, \bibinfo {author}
  {\bibfnamefont {C.}~\bibnamefont {Bradac}}, \bibinfo {author} {\bibfnamefont
  {I.}~\bibnamefont {Aharonovich}},\ and\ \bibinfo {author} {\bibfnamefont
  {V.}~\bibnamefont {Dyakonov}},\ }\bibfield  {title} {\bibinfo {title} {Spin
  defects in {hBN} as promising temperature, pressure and magnetic field
  quantum sensors},\ }\href@noop {} {\bibfield  {journal} {\bibinfo  {journal}
  {Nature communications}\ }\textbf {\bibinfo {volume} {12}},\ \bibinfo {pages}
  {4480} (\bibinfo {year} {2021}{\natexlab{b}})}\BibitemShut {NoStop}%
\bibitem [{\citenamefont {Liu}\ \emph {et~al.}(2021)\citenamefont {Liu},
  \citenamefont {Li}, \citenamefont {Yang}, \citenamefont {Yu}, \citenamefont
  {Meng}, \citenamefont {Wang}, \citenamefont {Li}, \citenamefont {Guo},
  \citenamefont {Yan}, \citenamefont {Li} \emph {et~al.}}]{liu2021temperature}%
  \BibitemOpen
  \bibfield  {author} {\bibinfo {author} {\bibfnamefont {W.}~\bibnamefont
  {Liu}}, \bibinfo {author} {\bibfnamefont {Z.-P.}\ \bibnamefont {Li}},
  \bibinfo {author} {\bibfnamefont {Y.-Z.}\ \bibnamefont {Yang}}, \bibinfo
  {author} {\bibfnamefont {S.}~\bibnamefont {Yu}}, \bibinfo {author}
  {\bibfnamefont {Y.}~\bibnamefont {Meng}}, \bibinfo {author} {\bibfnamefont
  {Z.-A.}\ \bibnamefont {Wang}}, \bibinfo {author} {\bibfnamefont {Z.-C.}\
  \bibnamefont {Li}}, \bibinfo {author} {\bibfnamefont {N.-J.}\ \bibnamefont
  {Guo}}, \bibinfo {author} {\bibfnamefont {F.-F.}\ \bibnamefont {Yan}},
  \bibinfo {author} {\bibfnamefont {Q.}~\bibnamefont {Li}}, \emph {et~al.},\
  }\bibfield  {title} {\bibinfo {title} {Temperature-dependent energy-level
  shifts of spin defects in hexagonal boron nitride},\ }\href@noop {}
  {\bibfield  {journal} {\bibinfo  {journal} {ACS Photonics}\ }\textbf
  {\bibinfo {volume} {8}},\ \bibinfo {pages} {1889} (\bibinfo {year}
  {2021})}\BibitemShut {NoStop}%
\bibitem [{\citenamefont {Yang}\ \emph {et~al.}(2022)\citenamefont {Yang},
  \citenamefont {Mendelson}, \citenamefont {Li}, \citenamefont {Gottscholl},
  \citenamefont {Scott}, \citenamefont {Kianinia}, \citenamefont {Dyakonov},
  \citenamefont {Toth},\ and\ \citenamefont {Aharonovich}}]{yang2022spin}%
  \BibitemOpen
  \bibfield  {author} {\bibinfo {author} {\bibfnamefont {T.}~\bibnamefont
  {Yang}}, \bibinfo {author} {\bibfnamefont {N.}~\bibnamefont {Mendelson}},
  \bibinfo {author} {\bibfnamefont {C.}~\bibnamefont {Li}}, \bibinfo {author}
  {\bibfnamefont {A.}~\bibnamefont {Gottscholl}}, \bibinfo {author}
  {\bibfnamefont {J.}~\bibnamefont {Scott}}, \bibinfo {author} {\bibfnamefont
  {M.}~\bibnamefont {Kianinia}}, \bibinfo {author} {\bibfnamefont
  {V.}~\bibnamefont {Dyakonov}}, \bibinfo {author} {\bibfnamefont
  {M.}~\bibnamefont {Toth}},\ and\ \bibinfo {author} {\bibfnamefont
  {I.}~\bibnamefont {Aharonovich}},\ }\bibfield  {title} {\bibinfo {title}
  {Spin defects in hexagonal boron nitride for strain sensing on nanopillar
  arrays},\ }\href@noop {} {\bibfield  {journal} {\bibinfo  {journal}
  {Nanoscale}\ }\textbf {\bibinfo {volume} {14}},\ \bibinfo {pages} {5239}
  (\bibinfo {year} {2022})}\BibitemShut {NoStop}%
\bibitem [{\citenamefont {Lyu}\ \emph {et~al.}(2022)\citenamefont {Lyu},
  \citenamefont {Tan}, \citenamefont {Wu}, \citenamefont {Zhang}, \citenamefont
  {Zhang}, \citenamefont {Mu}, \citenamefont {Z{\'u}{\~n}iga-P{\'e}rez},
  \citenamefont {Cai},\ and\ \citenamefont {Gao}}]{lyu2022strain}%
  \BibitemOpen
  \bibfield  {author} {\bibinfo {author} {\bibfnamefont {X.}~\bibnamefont
  {Lyu}}, \bibinfo {author} {\bibfnamefont {Q.}~\bibnamefont {Tan}}, \bibinfo
  {author} {\bibfnamefont {L.}~\bibnamefont {Wu}}, \bibinfo {author}
  {\bibfnamefont {C.}~\bibnamefont {Zhang}}, \bibinfo {author} {\bibfnamefont
  {Z.}~\bibnamefont {Zhang}}, \bibinfo {author} {\bibfnamefont
  {Z.}~\bibnamefont {Mu}}, \bibinfo {author} {\bibfnamefont {J.}~\bibnamefont
  {Z{\'u}{\~n}iga-P{\'e}rez}}, \bibinfo {author} {\bibfnamefont
  {H.}~\bibnamefont {Cai}},\ and\ \bibinfo {author} {\bibfnamefont
  {W.}~\bibnamefont {Gao}},\ }\bibfield  {title} {\bibinfo {title} {Strain
  quantum sensing with spin defects in hexagonal boron nitride},\ }\href@noop
  {} {\bibfield  {journal} {\bibinfo  {journal} {Nano Letters}\ }\textbf
  {\bibinfo {volume} {22}},\ \bibinfo {pages} {6553} (\bibinfo {year}
  {2022})}\BibitemShut {NoStop}%
\bibitem [{\citenamefont {Gao}\ \emph {et~al.}(2022)\citenamefont {Gao},
  \citenamefont {Vaidya}, \citenamefont {Li}, \citenamefont {Ju}, \citenamefont
  {Jiang}, \citenamefont {Xu}, \citenamefont {Allcca}, \citenamefont {Shen},
  \citenamefont {Taniguchi}, \citenamefont {Watanabe} \emph
  {et~al.}}]{gao2022nuclear}%
  \BibitemOpen
  \bibfield  {author} {\bibinfo {author} {\bibfnamefont {X.}~\bibnamefont
  {Gao}}, \bibinfo {author} {\bibfnamefont {S.}~\bibnamefont {Vaidya}},
  \bibinfo {author} {\bibfnamefont {K.}~\bibnamefont {Li}}, \bibinfo {author}
  {\bibfnamefont {P.}~\bibnamefont {Ju}}, \bibinfo {author} {\bibfnamefont
  {B.}~\bibnamefont {Jiang}}, \bibinfo {author} {\bibfnamefont
  {Z.}~\bibnamefont {Xu}}, \bibinfo {author} {\bibfnamefont {A.~E.~L.}\
  \bibnamefont {Allcca}}, \bibinfo {author} {\bibfnamefont {K.}~\bibnamefont
  {Shen}}, \bibinfo {author} {\bibfnamefont {T.}~\bibnamefont {Taniguchi}},
  \bibinfo {author} {\bibfnamefont {K.}~\bibnamefont {Watanabe}}, \emph
  {et~al.},\ }\bibfield  {title} {\bibinfo {title} {Nuclear spin polarization
  and control in hexagonal boron nitride},\ }\href@noop {} {\bibfield
  {journal} {\bibinfo  {journal} {Nature Materials}\ }\textbf {\bibinfo
  {volume} {21}},\ \bibinfo {pages} {1024} (\bibinfo {year}
  {2022})}\BibitemShut {NoStop}%
\bibitem [{\citenamefont {Thomas}(2015)}]{thomas2015breathing}%
  \BibitemOpen
  \bibfield  {author} {\bibinfo {author} {\bibfnamefont {D.~D.}\ \bibnamefont
  {Thomas}},\ }\bibfield  {title} {\bibinfo {title} {Breathing new life into
  nitric oxide signaling: a brief overview of the interplay between oxygen and
  nitric oxide},\ }\href@noop {} {\bibfield  {journal} {\bibinfo  {journal}
  {Redox biology}\ }\textbf {\bibinfo {volume} {5}},\ \bibinfo {pages} {225}
  (\bibinfo {year} {2015})}\BibitemShut {NoStop}%
\bibitem [{\citenamefont {Bogdan}(2015)}]{bogdan2015nitric}%
  \BibitemOpen
  \bibfield  {author} {\bibinfo {author} {\bibfnamefont {C.}~\bibnamefont
  {Bogdan}},\ }\bibfield  {title} {\bibinfo {title} {Nitric oxide synthase in
  innate and adaptive immunity: an update},\ }\href@noop {} {\bibfield
  {journal} {\bibinfo  {journal} {Trends in immunology}\ }\textbf {\bibinfo
  {volume} {36}},\ \bibinfo {pages} {161} (\bibinfo {year} {2015})}\BibitemShut
  {NoStop}%
\bibitem [{\citenamefont {Griendling}\ \emph {et~al.}(2016)\citenamefont
  {Griendling}, \citenamefont {Touyz}, \citenamefont {Zweier}, \citenamefont
  {Dikalov}, \citenamefont {Chilian}, \citenamefont {Chen}, \citenamefont
  {Harrison},\ and\ \citenamefont {Bhatnagar}}]{griendling2016measurement}%
  \BibitemOpen
  \bibfield  {author} {\bibinfo {author} {\bibfnamefont {K.~K.}\ \bibnamefont
  {Griendling}}, \bibinfo {author} {\bibfnamefont {R.~M.}\ \bibnamefont
  {Touyz}}, \bibinfo {author} {\bibfnamefont {J.~L.}\ \bibnamefont {Zweier}},
  \bibinfo {author} {\bibfnamefont {S.}~\bibnamefont {Dikalov}}, \bibinfo
  {author} {\bibfnamefont {W.}~\bibnamefont {Chilian}}, \bibinfo {author}
  {\bibfnamefont {Y.-R.}\ \bibnamefont {Chen}}, \bibinfo {author}
  {\bibfnamefont {D.~G.}\ \bibnamefont {Harrison}},\ and\ \bibinfo {author}
  {\bibfnamefont {A.}~\bibnamefont {Bhatnagar}},\ }\bibfield  {title} {\bibinfo
  {title} {Measurement of reactive oxygen species, reactive nitrogen species,
  and redox-dependent signaling in the cardiovascular system: a scientific
  statement from the american heart association},\ }\href@noop {} {\bibfield
  {journal} {\bibinfo  {journal} {Circulation research}\ }\textbf {\bibinfo
  {volume} {119}},\ \bibinfo {pages} {e39} (\bibinfo {year}
  {2016})}\BibitemShut {NoStop}%
\bibitem [{\citenamefont {Weinmann}\ \emph {et~al.}(1984)\citenamefont
  {Weinmann}, \citenamefont {Brasch}, \citenamefont {Press},\ and\
  \citenamefont {Wesbey}}]{weinmann1984characteristics}%
  \BibitemOpen
  \bibfield  {author} {\bibinfo {author} {\bibfnamefont {H.-J.}\ \bibnamefont
  {Weinmann}}, \bibinfo {author} {\bibfnamefont {R.~C.}\ \bibnamefont
  {Brasch}}, \bibinfo {author} {\bibfnamefont {W.-R.}\ \bibnamefont {Press}},\
  and\ \bibinfo {author} {\bibfnamefont {G.~E.}\ \bibnamefont {Wesbey}},\
  }\bibfield  {title} {\bibinfo {title} {Characteristics of gadolinium-{DTPA}
  complex: a potential {NMR} contrast agent},\ }\href@noop {} {\bibfield
  {journal} {\bibinfo  {journal} {American journal of roentgenology}\ }\textbf
  {\bibinfo {volume} {142}},\ \bibinfo {pages} {619} (\bibinfo {year}
  {1984})}\BibitemShut {NoStop}%
\bibitem [{\citenamefont {Chan}\ and\ \citenamefont
  {Wong}(2007)}]{chan2007small}%
  \BibitemOpen
  \bibfield  {author} {\bibinfo {author} {\bibfnamefont {K.~W.-Y.}\
  \bibnamefont {Chan}}\ and\ \bibinfo {author} {\bibfnamefont {W.-T.}\
  \bibnamefont {Wong}},\ }\bibfield  {title} {\bibinfo {title} {Small molecular
  gadolinium (iii) complexes as mri contrast agents for diagnostic imaging},\
  }\href@noop {} {\bibfield  {journal} {\bibinfo  {journal} {Coordination
  Chemistry Reviews}\ }\textbf {\bibinfo {volume} {251}},\ \bibinfo {pages}
  {2428} (\bibinfo {year} {2007})}\BibitemShut {NoStop}%
\bibitem [{\citenamefont {Tweedle}(2021)}]{doi:10.1148/radiol.2021210957}%
  \BibitemOpen
  \bibfield  {author} {\bibinfo {author} {\bibfnamefont {M.~F.}\ \bibnamefont
  {Tweedle}},\ }\bibfield  {title} {\bibinfo {title} {Gadolinium retention in
  human brain, bone, and skin},\ }\href
  {https://doi.org/10.1148/radiol.2021210957} {\bibfield  {journal} {\bibinfo
  {journal} {Radiology}\ }\textbf {\bibinfo {volume} {300}},\ \bibinfo {pages}
  {570} (\bibinfo {year} {2021})}\BibitemShut {NoStop}%
\bibitem [{\citenamefont {Dos~Santos}\ \emph {et~al.}(2022)\citenamefont
  {Dos~Santos}, \citenamefont {Parambathu}, \citenamefont {Fraenza},
  \citenamefont {Walsh}, \citenamefont {Greenbaum}, \citenamefont {Chapman},
  \citenamefont {Asthagiri},\ and\ \citenamefont {Singer}}]{dos2022thermal}%
  \BibitemOpen
  \bibfield  {author} {\bibinfo {author} {\bibfnamefont {T.~J.~P.}\
  \bibnamefont {Dos~Santos}}, \bibinfo {author} {\bibfnamefont {A.~V.}\
  \bibnamefont {Parambathu}}, \bibinfo {author} {\bibfnamefont {C.~C.}\
  \bibnamefont {Fraenza}}, \bibinfo {author} {\bibfnamefont {C.}~\bibnamefont
  {Walsh}}, \bibinfo {author} {\bibfnamefont {S.~G.}\ \bibnamefont
  {Greenbaum}}, \bibinfo {author} {\bibfnamefont {W.~G.}\ \bibnamefont
  {Chapman}}, \bibinfo {author} {\bibfnamefont {D.}~\bibnamefont {Asthagiri}},\
  and\ \bibinfo {author} {\bibfnamefont {P.~M.}\ \bibnamefont {Singer}},\
  }\bibfield  {title} {\bibinfo {title} {Thermal and concentration effects on
  $^1${H} {NMR} relaxation of {Gd}$^{3+}$-aqua using {MD} simulations and
  measurements},\ }\href@noop {} {\bibfield  {journal} {\bibinfo  {journal}
  {Physical Chemistry Chemical Physics}\ }\textbf {\bibinfo {volume} {24}},\
  \bibinfo {pages} {27964} (\bibinfo {year} {2022})}\BibitemShut {NoStop}%
\bibitem [{\citenamefont {Ma}\ \emph {et~al.}(2021)\citenamefont {Ma},
  \citenamefont {Azad}, \citenamefont {Dharmasivam}, \citenamefont
  {Richardson}, \citenamefont {Quinn}, \citenamefont {Feng}, \citenamefont
  {Pountney}, \citenamefont {Tonissen}, \citenamefont {Mellick}, \citenamefont
  {Yanatori} \emph {et~al.}}]{ma2021parkinson}%
  \BibitemOpen
  \bibfield  {author} {\bibinfo {author} {\bibfnamefont {L.}~\bibnamefont
  {Ma}}, \bibinfo {author} {\bibfnamefont {M.~G.}\ \bibnamefont {Azad}},
  \bibinfo {author} {\bibfnamefont {M.}~\bibnamefont {Dharmasivam}}, \bibinfo
  {author} {\bibfnamefont {V.}~\bibnamefont {Richardson}}, \bibinfo {author}
  {\bibfnamefont {R.}~\bibnamefont {Quinn}}, \bibinfo {author} {\bibfnamefont
  {Y.}~\bibnamefont {Feng}}, \bibinfo {author} {\bibfnamefont {D.}~\bibnamefont
  {Pountney}}, \bibinfo {author} {\bibfnamefont {K.}~\bibnamefont {Tonissen}},
  \bibinfo {author} {\bibfnamefont {G.}~\bibnamefont {Mellick}}, \bibinfo
  {author} {\bibfnamefont {I.}~\bibnamefont {Yanatori}}, \emph {et~al.},\
  }\bibfield  {title} {\bibinfo {title} {Parkinson's disease: alterations in
  iron and redox biology as a key to unlock therapeutic strategies},\
  }\href@noop {} {\bibfield  {journal} {\bibinfo  {journal} {Redox Biology}\
  }\textbf {\bibinfo {volume} {41}},\ \bibinfo {pages} {101896} (\bibinfo
  {year} {2021})}\BibitemShut {NoStop}%
\bibitem [{\citenamefont {Samrot}\ \emph {et~al.}(2021)\citenamefont {Samrot},
  \citenamefont {Sahithya}, \citenamefont {Selvarani}, \citenamefont
  {Purayil},\ and\ \citenamefont {Ponnaiah}}]{samrot2021review}%
  \BibitemOpen
  \bibfield  {author} {\bibinfo {author} {\bibfnamefont {A.~V.}\ \bibnamefont
  {Samrot}}, \bibinfo {author} {\bibfnamefont {C.~S.}\ \bibnamefont
  {Sahithya}}, \bibinfo {author} {\bibfnamefont {J.}~\bibnamefont {Selvarani}},
  \bibinfo {author} {\bibfnamefont {S.~K.}\ \bibnamefont {Purayil}},\ and\
  \bibinfo {author} {\bibfnamefont {P.}~\bibnamefont {Ponnaiah}},\ }\bibfield
  {title} {\bibinfo {title} {A review on synthesis, characterization and
  potential biological applications of superparamagnetic iron oxide
  nanoparticles},\ }\href@noop {} {\bibfield  {journal} {\bibinfo  {journal}
  {Current Research in Green and Sustainable Chemistry}\ }\textbf {\bibinfo
  {volume} {4}},\ \bibinfo {pages} {100042} (\bibinfo {year}
  {2021})}\BibitemShut {NoStop}%
\bibitem [{\citenamefont {Steinert}\ \emph {et~al.}(2013)\citenamefont
  {Steinert}, \citenamefont {Ziem}, \citenamefont {Hall}, \citenamefont
  {Zappe}, \citenamefont {Schweikert}, \citenamefont {G{\"o}tz}, \citenamefont
  {Aird}, \citenamefont {Balasubramanian}, \citenamefont {Hollenberg},\ and\
  \citenamefont {Wrachtrup}}]{steinert2013magnetic}%
  \BibitemOpen
  \bibfield  {author} {\bibinfo {author} {\bibfnamefont {S.}~\bibnamefont
  {Steinert}}, \bibinfo {author} {\bibfnamefont {F.}~\bibnamefont {Ziem}},
  \bibinfo {author} {\bibfnamefont {L.}~\bibnamefont {Hall}}, \bibinfo {author}
  {\bibfnamefont {A.}~\bibnamefont {Zappe}}, \bibinfo {author} {\bibfnamefont
  {M.}~\bibnamefont {Schweikert}}, \bibinfo {author} {\bibfnamefont
  {N.}~\bibnamefont {G{\"o}tz}}, \bibinfo {author} {\bibfnamefont
  {A.}~\bibnamefont {Aird}}, \bibinfo {author} {\bibfnamefont {G.}~\bibnamefont
  {Balasubramanian}}, \bibinfo {author} {\bibfnamefont {L.}~\bibnamefont
  {Hollenberg}},\ and\ \bibinfo {author} {\bibfnamefont {J.}~\bibnamefont
  {Wrachtrup}},\ }\bibfield  {title} {\bibinfo {title} {Magnetic spin imaging
  under ambient conditions with sub-cellular resolution},\ }\href@noop {}
  {\bibfield  {journal} {\bibinfo  {journal} {Nature communications}\ }\textbf
  {\bibinfo {volume} {4}},\ \bibinfo {pages} {1607} (\bibinfo {year}
  {2013})}\BibitemShut {NoStop}%
\bibitem [{\citenamefont {Ziem}\ \emph {et~al.}(2013)\citenamefont {Ziem},
  \citenamefont {Gotz}, \citenamefont {Zappe}, \citenamefont {Steinert},\ and\
  \citenamefont {Wrachtrup}}]{ziem2013highly}%
  \BibitemOpen
  \bibfield  {author} {\bibinfo {author} {\bibfnamefont {F.~C.}\ \bibnamefont
  {Ziem}}, \bibinfo {author} {\bibfnamefont {N.~S.}\ \bibnamefont {Gotz}},
  \bibinfo {author} {\bibfnamefont {A.}~\bibnamefont {Zappe}}, \bibinfo
  {author} {\bibfnamefont {S.}~\bibnamefont {Steinert}},\ and\ \bibinfo
  {author} {\bibfnamefont {J.}~\bibnamefont {Wrachtrup}},\ }\bibfield  {title}
  {\bibinfo {title} {Highly sensitive detection of physiological spins in a
  microfluidic device},\ }\href@noop {} {\bibfield  {journal} {\bibinfo
  {journal} {Nano letters}\ }\textbf {\bibinfo {volume} {13}},\ \bibinfo
  {pages} {4093} (\bibinfo {year} {2013})}\BibitemShut {NoStop}%
\bibitem [{\citenamefont {Ermakova}\ \emph {et~al.}(2013)\citenamefont
  {Ermakova}, \citenamefont {Pramanik}, \citenamefont {Cai}, \citenamefont
  {Algara-Siller}, \citenamefont {Kaiser}, \citenamefont {Weil}, \citenamefont
  {Tzeng}, \citenamefont {Chang}, \citenamefont {McGuinness}, \citenamefont
  {Plenio} \emph {et~al.}}]{ermakova2013detection}%
  \BibitemOpen
  \bibfield  {author} {\bibinfo {author} {\bibfnamefont {A.}~\bibnamefont
  {Ermakova}}, \bibinfo {author} {\bibfnamefont {G.}~\bibnamefont {Pramanik}},
  \bibinfo {author} {\bibfnamefont {J.-M.}\ \bibnamefont {Cai}}, \bibinfo
  {author} {\bibfnamefont {G.}~\bibnamefont {Algara-Siller}}, \bibinfo {author}
  {\bibfnamefont {U.}~\bibnamefont {Kaiser}}, \bibinfo {author} {\bibfnamefont
  {T.}~\bibnamefont {Weil}}, \bibinfo {author} {\bibfnamefont {Y.-K.}\
  \bibnamefont {Tzeng}}, \bibinfo {author} {\bibfnamefont {H.-C.}\ \bibnamefont
  {Chang}}, \bibinfo {author} {\bibfnamefont {L.}~\bibnamefont {McGuinness}},
  \bibinfo {author} {\bibfnamefont {M.~B.}\ \bibnamefont {Plenio}}, \emph
  {et~al.},\ }\bibfield  {title} {\bibinfo {title} {Detection of a few
  metallo-protein molecules using color centers in nanodiamonds},\ }\href@noop
  {} {\bibfield  {journal} {\bibinfo  {journal} {Nano letters}\ }\textbf
  {\bibinfo {volume} {13}},\ \bibinfo {pages} {3305} (\bibinfo {year}
  {2013})}\BibitemShut {NoStop}%
\bibitem [{\citenamefont {Shi}\ \emph {et~al.}(2015)\citenamefont {Shi},
  \citenamefont {Zhang}, \citenamefont {Wang}, \citenamefont {Sun},
  \citenamefont {Wang}, \citenamefont {Rong}, \citenamefont {Chen},
  \citenamefont {Ju}, \citenamefont {Reinhard}, \citenamefont {Chen} \emph
  {et~al.}}]{shi2015single}%
  \BibitemOpen
  \bibfield  {author} {\bibinfo {author} {\bibfnamefont {F.}~\bibnamefont
  {Shi}}, \bibinfo {author} {\bibfnamefont {Q.}~\bibnamefont {Zhang}}, \bibinfo
  {author} {\bibfnamefont {P.}~\bibnamefont {Wang}}, \bibinfo {author}
  {\bibfnamefont {H.}~\bibnamefont {Sun}}, \bibinfo {author} {\bibfnamefont
  {J.}~\bibnamefont {Wang}}, \bibinfo {author} {\bibfnamefont {X.}~\bibnamefont
  {Rong}}, \bibinfo {author} {\bibfnamefont {M.}~\bibnamefont {Chen}}, \bibinfo
  {author} {\bibfnamefont {C.}~\bibnamefont {Ju}}, \bibinfo {author}
  {\bibfnamefont {F.}~\bibnamefont {Reinhard}}, \bibinfo {author}
  {\bibfnamefont {H.}~\bibnamefont {Chen}}, \emph {et~al.},\ }\bibfield
  {title} {\bibinfo {title} {Single-protein spin resonance spectroscopy under
  ambient conditions},\ }\href@noop {} {\bibfield  {journal} {\bibinfo
  {journal} {Science}\ }\textbf {\bibinfo {volume} {347}},\ \bibinfo {pages}
  {1135} (\bibinfo {year} {2015})}\BibitemShut {NoStop}%
\bibitem [{\citenamefont {Simpson}\ \emph {et~al.}(2017)\citenamefont
  {Simpson}, \citenamefont {Ryan}, \citenamefont {Hall}, \citenamefont
  {Panchenko}, \citenamefont {Drew}, \citenamefont {Petrou}, \citenamefont
  {Donnelly}, \citenamefont {Mulvaney},\ and\ \citenamefont
  {Hollenberg}}]{simpson2017electron}%
  \BibitemOpen
  \bibfield  {author} {\bibinfo {author} {\bibfnamefont {D.~A.}\ \bibnamefont
  {Simpson}}, \bibinfo {author} {\bibfnamefont {R.~G.}\ \bibnamefont {Ryan}},
  \bibinfo {author} {\bibfnamefont {L.~T.}\ \bibnamefont {Hall}}, \bibinfo
  {author} {\bibfnamefont {E.}~\bibnamefont {Panchenko}}, \bibinfo {author}
  {\bibfnamefont {S.~C.}\ \bibnamefont {Drew}}, \bibinfo {author}
  {\bibfnamefont {S.}~\bibnamefont {Petrou}}, \bibinfo {author} {\bibfnamefont
  {P.~S.}\ \bibnamefont {Donnelly}}, \bibinfo {author} {\bibfnamefont
  {P.}~\bibnamefont {Mulvaney}},\ and\ \bibinfo {author} {\bibfnamefont
  {L.~C.}\ \bibnamefont {Hollenberg}},\ }\bibfield  {title} {\bibinfo {title}
  {Electron paramagnetic resonance microscopy using spins in diamond under
  ambient conditions},\ }\href@noop {} {\bibfield  {journal} {\bibinfo
  {journal} {Nature Communications}\ }\textbf {\bibinfo {volume} {8}},\
  \bibinfo {pages} {458} (\bibinfo {year} {2017})}\BibitemShut {NoStop}%
\bibitem [{\citenamefont {Gorrini}\ \emph {et~al.}(2019)\citenamefont
  {Gorrini}, \citenamefont {Giri}, \citenamefont {Avalos}, \citenamefont
  {Tambalo}, \citenamefont {Mannucci}, \citenamefont {Basso}, \citenamefont
  {Bazzanella}, \citenamefont {Dorigoni}, \citenamefont {Cazzanelli},
  \citenamefont {Marzola} \emph {et~al.}}]{gorrini2019fast}%
  \BibitemOpen
  \bibfield  {author} {\bibinfo {author} {\bibfnamefont {F.}~\bibnamefont
  {Gorrini}}, \bibinfo {author} {\bibfnamefont {R.}~\bibnamefont {Giri}},
  \bibinfo {author} {\bibfnamefont {C.}~\bibnamefont {Avalos}}, \bibinfo
  {author} {\bibfnamefont {S.}~\bibnamefont {Tambalo}}, \bibinfo {author}
  {\bibfnamefont {S.}~\bibnamefont {Mannucci}}, \bibinfo {author}
  {\bibfnamefont {L.}~\bibnamefont {Basso}}, \bibinfo {author} {\bibfnamefont
  {N.}~\bibnamefont {Bazzanella}}, \bibinfo {author} {\bibfnamefont
  {C.}~\bibnamefont {Dorigoni}}, \bibinfo {author} {\bibfnamefont
  {M.}~\bibnamefont {Cazzanelli}}, \bibinfo {author} {\bibfnamefont
  {P.}~\bibnamefont {Marzola}}, \emph {et~al.},\ }\bibfield  {title} {\bibinfo
  {title} {Fast and sensitive detection of paramagnetic species using coupled
  charge and spin dynamics in strongly fluorescent nanodiamonds},\ }\href@noop
  {} {\bibfield  {journal} {\bibinfo  {journal} {ACS applied materials \&
  interfaces}\ }\textbf {\bibinfo {volume} {11}},\ \bibinfo {pages} {24412}
  (\bibinfo {year} {2019})}\BibitemShut {NoStop}%
\bibitem [{\citenamefont {Radu}\ \emph {et~al.}(2019)\citenamefont {Radu},
  \citenamefont {Price}, \citenamefont {Levett}, \citenamefont {Narayanasamy},
  \citenamefont {Bateman-Price}, \citenamefont {Wilson},\ and\ \citenamefont
  {Mather}}]{radu2019dynamic}%
  \BibitemOpen
  \bibfield  {author} {\bibinfo {author} {\bibfnamefont {V.}~\bibnamefont
  {Radu}}, \bibinfo {author} {\bibfnamefont {J.~C.}\ \bibnamefont {Price}},
  \bibinfo {author} {\bibfnamefont {S.~J.}\ \bibnamefont {Levett}}, \bibinfo
  {author} {\bibfnamefont {K.~K.}\ \bibnamefont {Narayanasamy}}, \bibinfo
  {author} {\bibfnamefont {T.~D.}\ \bibnamefont {Bateman-Price}}, \bibinfo
  {author} {\bibfnamefont {P.~B.}\ \bibnamefont {Wilson}},\ and\ \bibinfo
  {author} {\bibfnamefont {M.~L.}\ \bibnamefont {Mather}},\ }\bibfield  {title}
  {\bibinfo {title} {Dynamic quantum sensing of paramagnetic species using
  nitrogen-vacancy centers in diamond},\ }\href@noop {} {\bibfield  {journal}
  {\bibinfo  {journal} {ACS sensors}\ }\textbf {\bibinfo {volume} {5}}
  (\bibinfo {year} {2019})}\BibitemShut {NoStop}%
\bibitem [{\citenamefont {Liu}\ \emph {et~al.}(2022)\citenamefont {Liu},
  \citenamefont {Guo}, \citenamefont {Yu}, \citenamefont {Meng}, \citenamefont
  {Li}, \citenamefont {Yang}, \citenamefont {Wang}, \citenamefont {Zeng},
  \citenamefont {Xie}, \citenamefont {Wang} \emph {et~al.}}]{liu2022spin}%
  \BibitemOpen
  \bibfield  {author} {\bibinfo {author} {\bibfnamefont {W.}~\bibnamefont
  {Liu}}, \bibinfo {author} {\bibfnamefont {N.-J.}\ \bibnamefont {Guo}},
  \bibinfo {author} {\bibfnamefont {S.}~\bibnamefont {Yu}}, \bibinfo {author}
  {\bibfnamefont {Y.}~\bibnamefont {Meng}}, \bibinfo {author} {\bibfnamefont
  {Z.}~\bibnamefont {Li}}, \bibinfo {author} {\bibfnamefont {Y.-Z.}\
  \bibnamefont {Yang}}, \bibinfo {author} {\bibfnamefont {Z.-A.}\ \bibnamefont
  {Wang}}, \bibinfo {author} {\bibfnamefont {X.-D.}\ \bibnamefont {Zeng}},
  \bibinfo {author} {\bibfnamefont {L.-K.}\ \bibnamefont {Xie}}, \bibinfo
  {author} {\bibfnamefont {J.-F.}\ \bibnamefont {Wang}}, \emph {et~al.},\
  }\bibfield  {title} {\bibinfo {title} {Spin-active defects in hexagonal boron
  nitride},\ }\href@noop {} {\bibfield  {journal} {\bibinfo  {journal}
  {Materials for Quantum Technology}\ }\textbf {\bibinfo {volume} {1}},\
  \bibinfo {pages} {032002} (\bibinfo {year} {2022})}\BibitemShut {NoStop}%
\bibitem [{\citenamefont {Xu}\ \emph {et~al.}(2023)\citenamefont {Xu},
  \citenamefont {Solanki}, \citenamefont {Sychev}, \citenamefont {Gao},
  \citenamefont {Peana}, \citenamefont {Baburin}, \citenamefont {Pagadala},
  \citenamefont {Martin}, \citenamefont {Chowdhury}, \citenamefont {Chen} \emph
  {et~al.}}]{xu2022greatly}%
  \BibitemOpen
  \bibfield  {author} {\bibinfo {author} {\bibfnamefont {X.}~\bibnamefont
  {Xu}}, \bibinfo {author} {\bibfnamefont {A.~B.}\ \bibnamefont {Solanki}},
  \bibinfo {author} {\bibfnamefont {D.}~\bibnamefont {Sychev}}, \bibinfo
  {author} {\bibfnamefont {X.}~\bibnamefont {Gao}}, \bibinfo {author}
  {\bibfnamefont {S.}~\bibnamefont {Peana}}, \bibinfo {author} {\bibfnamefont
  {A.~S.}\ \bibnamefont {Baburin}}, \bibinfo {author} {\bibfnamefont
  {K.}~\bibnamefont {Pagadala}}, \bibinfo {author} {\bibfnamefont {Z.~O.}\
  \bibnamefont {Martin}}, \bibinfo {author} {\bibfnamefont {S.~N.}\
  \bibnamefont {Chowdhury}}, \bibinfo {author} {\bibfnamefont {Y.~P.}\
  \bibnamefont {Chen}}, \emph {et~al.},\ }\bibfield  {title} {\bibinfo {title}
  {Greatly enhanced emission from spin defects in hexagonal boron nitride
  enabled by a low-loss plasmonic nanocavity},\ }\href@noop {} {\bibfield
  {journal} {\bibinfo  {journal} {Nano Letters}\ }\textbf {\bibinfo {volume}
  {23}},\ \bibinfo {pages} {25–33} (\bibinfo {year} {2023})}\BibitemShut
  {NoStop}%
\bibitem [{\citenamefont {Iv{\'a}dy}\ \emph {et~al.}(2020)\citenamefont
  {Iv{\'a}dy}, \citenamefont {Barcza}, \citenamefont {Thiering}, \citenamefont
  {Li}, \citenamefont {Hamdi}, \citenamefont {Chou}, \citenamefont {Legeza},\
  and\ \citenamefont {Gali}}]{ivady2020ab}%
  \BibitemOpen
  \bibfield  {author} {\bibinfo {author} {\bibfnamefont {V.}~\bibnamefont
  {Iv{\'a}dy}}, \bibinfo {author} {\bibfnamefont {G.}~\bibnamefont {Barcza}},
  \bibinfo {author} {\bibfnamefont {G.}~\bibnamefont {Thiering}}, \bibinfo
  {author} {\bibfnamefont {S.}~\bibnamefont {Li}}, \bibinfo {author}
  {\bibfnamefont {H.}~\bibnamefont {Hamdi}}, \bibinfo {author} {\bibfnamefont
  {J.-P.}\ \bibnamefont {Chou}}, \bibinfo {author} {\bibfnamefont
  {{\"O}.}~\bibnamefont {Legeza}},\ and\ \bibinfo {author} {\bibfnamefont
  {A.}~\bibnamefont {Gali}},\ }\bibfield  {title} {\bibinfo {title} {Ab initio
  theory of the negatively charged boron vacancy qubit in hexagonal boron
  nitride},\ }\href@noop {} {\bibfield  {journal} {\bibinfo  {journal} {npj
  Computational Materials}\ }\textbf {\bibinfo {volume} {6}},\ \bibinfo {pages}
  {41} (\bibinfo {year} {2020})}\BibitemShut {NoStop}%
\bibitem [{\citenamefont {Mathur}\ \emph {et~al.}(2022)\citenamefont {Mathur},
  \citenamefont {Mukherjee}, \citenamefont {Gao}, \citenamefont {Luo},
  \citenamefont {McCullian}, \citenamefont {Li}, \citenamefont {Vamivakas},\
  and\ \citenamefont {Fuchs}}]{mathur2022excited}%
  \BibitemOpen
  \bibfield  {author} {\bibinfo {author} {\bibfnamefont {N.}~\bibnamefont
  {Mathur}}, \bibinfo {author} {\bibfnamefont {A.}~\bibnamefont {Mukherjee}},
  \bibinfo {author} {\bibfnamefont {X.}~\bibnamefont {Gao}}, \bibinfo {author}
  {\bibfnamefont {J.}~\bibnamefont {Luo}}, \bibinfo {author} {\bibfnamefont
  {B.~A.}\ \bibnamefont {McCullian}}, \bibinfo {author} {\bibfnamefont
  {T.}~\bibnamefont {Li}}, \bibinfo {author} {\bibfnamefont {A.~N.}\
  \bibnamefont {Vamivakas}},\ and\ \bibinfo {author} {\bibfnamefont {G.~D.}\
  \bibnamefont {Fuchs}},\ }\bibfield  {title} {\bibinfo {title} {Excited-state
  spin-resonance spectroscopy of {$V_B^-$} defect centers in hexagonal boron
  nitride},\ }\href@noop {} {\bibfield  {journal} {\bibinfo  {journal} {Nature
  Communications}\ }\textbf {\bibinfo {volume} {13}},\ \bibinfo {pages} {3233}
  (\bibinfo {year} {2022})}\BibitemShut {NoStop}%
\bibitem [{\citenamefont {Baber}\ \emph {et~al.}(2021)\citenamefont {Baber},
  \citenamefont {Malein}, \citenamefont {Khatri}, \citenamefont {Keatley},
  \citenamefont {Guo}, \citenamefont {Withers}, \citenamefont {Ramsay},\ and\
  \citenamefont {Luxmoore}}]{baber2021excited}%
  \BibitemOpen
  \bibfield  {author} {\bibinfo {author} {\bibfnamefont {S.}~\bibnamefont
  {Baber}}, \bibinfo {author} {\bibfnamefont {R.~N.~E.}\ \bibnamefont
  {Malein}}, \bibinfo {author} {\bibfnamefont {P.}~\bibnamefont {Khatri}},
  \bibinfo {author} {\bibfnamefont {P.~S.}\ \bibnamefont {Keatley}}, \bibinfo
  {author} {\bibfnamefont {S.}~\bibnamefont {Guo}}, \bibinfo {author}
  {\bibfnamefont {F.}~\bibnamefont {Withers}}, \bibinfo {author} {\bibfnamefont
  {A.~J.}\ \bibnamefont {Ramsay}},\ and\ \bibinfo {author} {\bibfnamefont
  {I.~J.}\ \bibnamefont {Luxmoore}},\ }\bibfield  {title} {\bibinfo {title}
  {Excited state spectroscopy of boron vacancy defects in hexagonal boron
  nitride using time-resolved optically detected magnetic resonance},\
  }\href@noop {} {\bibfield  {journal} {\bibinfo  {journal} {Nano Letters}\
  }\textbf {\bibinfo {volume} {22}},\ \bibinfo {pages} {461} (\bibinfo {year}
  {2021})}\BibitemShut {NoStop}%
\bibitem [{\citenamefont {Mu}\ \emph {et~al.}(2022)\citenamefont {Mu},
  \citenamefont {Cai}, \citenamefont {Chen}, \citenamefont {Kenny},
  \citenamefont {Jiang}, \citenamefont {Ru}, \citenamefont {Lyu}, \citenamefont
  {Koh}, \citenamefont {Liu}, \citenamefont {Aharonovich} \emph
  {et~al.}}]{mu2022excited}%
  \BibitemOpen
  \bibfield  {author} {\bibinfo {author} {\bibfnamefont {Z.}~\bibnamefont
  {Mu}}, \bibinfo {author} {\bibfnamefont {H.}~\bibnamefont {Cai}}, \bibinfo
  {author} {\bibfnamefont {D.}~\bibnamefont {Chen}}, \bibinfo {author}
  {\bibfnamefont {J.}~\bibnamefont {Kenny}}, \bibinfo {author} {\bibfnamefont
  {Z.}~\bibnamefont {Jiang}}, \bibinfo {author} {\bibfnamefont
  {S.}~\bibnamefont {Ru}}, \bibinfo {author} {\bibfnamefont {X.}~\bibnamefont
  {Lyu}}, \bibinfo {author} {\bibfnamefont {T.~S.}\ \bibnamefont {Koh}},
  \bibinfo {author} {\bibfnamefont {X.}~\bibnamefont {Liu}}, \bibinfo {author}
  {\bibfnamefont {I.}~\bibnamefont {Aharonovich}}, \emph {et~al.},\ }\bibfield
  {title} {\bibinfo {title} {Excited-state optically detected magnetic
  resonance of spin defects in hexagonal boron nitride},\ }\href@noop {}
  {\bibfield  {journal} {\bibinfo  {journal} {Physical Review Letters}\
  }\textbf {\bibinfo {volume} {128}},\ \bibinfo {pages} {216402} (\bibinfo
  {year} {2022})}\BibitemShut {NoStop}%
\bibitem [{\citenamefont {Yu}\ \emph {et~al.}(2022)\citenamefont {Yu},
  \citenamefont {Sun}, \citenamefont {Wang}, \citenamefont {Zhang},
  \citenamefont {Ye}, \citenamefont {Zhou}, \citenamefont {Liu}, \citenamefont
  {Wang}, \citenamefont {Shi}, \citenamefont {Wang} \emph
  {et~al.}}]{yu2022excited}%
  \BibitemOpen
  \bibfield  {author} {\bibinfo {author} {\bibfnamefont {P.}~\bibnamefont
  {Yu}}, \bibinfo {author} {\bibfnamefont {H.}~\bibnamefont {Sun}}, \bibinfo
  {author} {\bibfnamefont {M.}~\bibnamefont {Wang}}, \bibinfo {author}
  {\bibfnamefont {T.}~\bibnamefont {Zhang}}, \bibinfo {author} {\bibfnamefont
  {X.}~\bibnamefont {Ye}}, \bibinfo {author} {\bibfnamefont {J.}~\bibnamefont
  {Zhou}}, \bibinfo {author} {\bibfnamefont {H.}~\bibnamefont {Liu}}, \bibinfo
  {author} {\bibfnamefont {C.-J.}\ \bibnamefont {Wang}}, \bibinfo {author}
  {\bibfnamefont {F.}~\bibnamefont {Shi}}, \bibinfo {author} {\bibfnamefont
  {Y.}~\bibnamefont {Wang}}, \emph {et~al.},\ }\bibfield  {title} {\bibinfo
  {title} {Excited-state spectroscopy of spin defects in hexagonal boron
  nitride},\ }\href@noop {} {\bibfield  {journal} {\bibinfo  {journal} {Nano
  Letters}\ }\textbf {\bibinfo {volume} {22}},\ \bibinfo {pages} {3545}
  (\bibinfo {year} {2022})}\BibitemShut {NoStop}%
\bibitem [{\citenamefont {Ziegler}\ \emph {et~al.}(2010)\citenamefont
  {Ziegler}, \citenamefont {Ziegler},\ and\ \citenamefont
  {Biersack}}]{ziegler2010srim}%
  \BibitemOpen
  \bibfield  {author} {\bibinfo {author} {\bibfnamefont {J.~F.}\ \bibnamefont
  {Ziegler}}, \bibinfo {author} {\bibfnamefont {M.~D.}\ \bibnamefont
  {Ziegler}},\ and\ \bibinfo {author} {\bibfnamefont {J.~P.}\ \bibnamefont
  {Biersack}},\ }\bibfield  {title} {\bibinfo {title} {{SRIM}--the stopping and
  range of ions in matter (2010)},\ }\href@noop {} {\bibfield  {journal}
  {\bibinfo  {journal} {Nuclear Instruments and Methods in Physics Research
  Section B: Beam Interactions with Materials and Atoms}\ }\textbf {\bibinfo
  {volume} {268}},\ \bibinfo {pages} {1818} (\bibinfo {year}
  {2010})}\BibitemShut {NoStop}%
\bibitem [{\citenamefont {Gong}\ \emph {et~al.}(2022)\citenamefont {Gong},
  \citenamefont {He}, \citenamefont {Gao}, \citenamefont {Ju}, \citenamefont
  {Liu}, \citenamefont {Ye}, \citenamefont {Henriksen}, \citenamefont {Li},\
  and\ \citenamefont {Zu}}]{gong2022coherent}%
  \BibitemOpen
  \bibfield  {author} {\bibinfo {author} {\bibfnamefont {R.}~\bibnamefont
  {Gong}}, \bibinfo {author} {\bibfnamefont {G.}~\bibnamefont {He}}, \bibinfo
  {author} {\bibfnamefont {X.}~\bibnamefont {Gao}}, \bibinfo {author}
  {\bibfnamefont {P.}~\bibnamefont {Ju}}, \bibinfo {author} {\bibfnamefont
  {Z.}~\bibnamefont {Liu}}, \bibinfo {author} {\bibfnamefont {B.}~\bibnamefont
  {Ye}}, \bibinfo {author} {\bibfnamefont {E.~A.}\ \bibnamefont {Henriksen}},
  \bibinfo {author} {\bibfnamefont {T.}~\bibnamefont {Li}},\ and\ \bibinfo
  {author} {\bibfnamefont {C.}~\bibnamefont {Zu}},\ }\bibfield  {title}
  {\bibinfo {title} {Coherent dynamics of strongly interacting electronic spin
  defects in hexagonal boron nitride},\ }\href@noop {} {\bibfield  {journal}
  {\bibinfo  {journal} {arXiv preprint arXiv:2210.11485}\ } (\bibinfo {year}
  {2022})}\BibitemShut {NoStop}%
\bibitem [{\citenamefont {Robertson}\ \emph {et~al.}(2023)\citenamefont
  {Robertson}, \citenamefont {Scholten}, \citenamefont {Singh}, \citenamefont
  {Healey}, \citenamefont {Meneses}, \citenamefont {Reineck}, \citenamefont
  {Abe}, \citenamefont {Ohshima}, \citenamefont {Kianinia}, \citenamefont
  {Aharonovich} \emph {et~al.}}]{robertson2023detection}%
  \BibitemOpen
  \bibfield  {author} {\bibinfo {author} {\bibfnamefont {I.~O.}\ \bibnamefont
  {Robertson}}, \bibinfo {author} {\bibfnamefont {S.~C.}\ \bibnamefont
  {Scholten}}, \bibinfo {author} {\bibfnamefont {P.}~\bibnamefont {Singh}},
  \bibinfo {author} {\bibfnamefont {A.~J.}\ \bibnamefont {Healey}}, \bibinfo
  {author} {\bibfnamefont {F.}~\bibnamefont {Meneses}}, \bibinfo {author}
  {\bibfnamefont {P.}~\bibnamefont {Reineck}}, \bibinfo {author} {\bibfnamefont
  {H.}~\bibnamefont {Abe}}, \bibinfo {author} {\bibfnamefont {T.}~\bibnamefont
  {Ohshima}}, \bibinfo {author} {\bibfnamefont {M.}~\bibnamefont {Kianinia}},
  \bibinfo {author} {\bibfnamefont {I.}~\bibnamefont {Aharonovich}}, \emph
  {et~al.},\ }\bibfield  {title} {\bibinfo {title} {Detection of paramagnetic
  spins with an ultrathin van der waals quantum sensor},\ }\href@noop {}
  {\bibfield  {journal} {\bibinfo  {journal} {arXiv preprint arXiv:2302.10560}\
  } (\bibinfo {year} {2023})}\BibitemShut {NoStop}%
\end{thebibliography}

\end{document}